\documentclass[preprint2]{aastex}

\usepackage{epsfig}

\newcommand\etal{{\it et al}}                                                  
\def\etal{{\it et al}}
\def\deg{^{\circ}}
\def\P3hat{{\mathaccent 94 P}_3}

\def\eg{{\it e.g.}}

\shorttitle{J. M. Rankin, & R. Ramachandran}
\shortauthors{Pulsar Emission Mapping}

\begin{document}

\title{Toward An Empirical Theory of Pulsar Emission VIII:  Subbeam 
Circulation and the Polarization-Modal Structure of Conal Beams}

\author{Joanna M. Rankin\footnote{On leave from Physics Department, 
A405 Cook Bldg., University of Vermont, Burlington, VT 05405 USA; 
email: rankin@physics.uvm.edu}}
\affil{Sterrenkundig Instituut `Anton Pannekoek', Universiteit van Amsterdam, 
1098 SJ Amsterdam NL}

\author{R. Ramachandran\footnote{Current address: Astronomy Department, 
601 Campbell Hall, University of California, Berkeley, CA 94720-3411 USA; 
email: ramach@astron.berkely.edu}}
\affil{Stichting ASTRON, Postbus 2, 7990 AA Dwingeloo, The Netherlands}


\begin{abstract}
The average polarization properties of conal single and double
profiles directly reflect the polarization-modal structure of the
emission beams which produce them.  Conal component pairs exhibit
large fractional linear polarization on their inside edges and
virtually complete depolarization on their outside edges; whereas
profiles resulting from sightline encounters with the outer conal 
edge are usually very depolarized.  The polarization-modal character 
of subbeam circulation produces conditions whereby both angular and
temporal averaging contribute to this polarization and depolarization.

These circumstances combine to require that the circulating subbeam
systems which produce conal beams entail paired PPM and SPM emission
elements which are offset from each other in both magnetic azimuth 
and magnetic colatitude.  Or, as rotating subbeam systems produce 
(on average) conal beams, one modal subcone has a little larger 
(or smaller) radius than the other.  However, these PPM and SPM 
``beamlets'' cannot be in azimuthal phase, because both alternately 
dominate the emission on the extreme outer edge of the conal beam. 
While this configuration can be deduced from the observations, 
simulation of this rotating, modal subbeam system reiterates these 
these conclusions.  These circumstances are also probably responsible, 
along with the usual wavelength dependence of emission height, for 
the observed spectral decline in aggregate polarization.

A clear delineation of the modal polarization topology of the conal
beam promises to address fundamental questions about the nature and
origin of this modal emission---and the modal parity at the outer 
beam edges is a fact of considerable significance.  The different 
angular dependences of the modal ``beamlets'' suggests that the 
polarization modes are generated via propagation effects.  This 
argument may prove much stronger if the modal emission is 
fundamentally only partially polarized.  Several theories now 
promise quantitative comparison with the observations.
\end{abstract}




%

\section*{The Outer-Edge Depolarization Phenomenon}

A humble fact about pulsar radio emission, which to our knowledge has
attracted virtually no notice or comment, is the following: The
extreme outer edges of virtually all conal component pairs are
prominently, and apparently accurately, depolarized.  Considerable
comment {\it has} been made regarding the obverse of this
circumstance---that is, to the effect that the highest levels of
fractional linear polarization are usually found on the inner edges of
conal components, indeed where it is sometimes nearly complete
(Manchester 1971; Morris \etal\ 1981).

Of course, longitudes corresponding to the outer edges of such conal
component pairs are also just where intervals of secondary
polarization-mode dominance are seen in individual pulses, as we know
from those well known stars whose profiles indicate a fairly central
sightline traverse through the conal beam: B0329+54 exhibits zones of
outer edge depolarization over a wide frequency range (Gould \& Lyne
1998; hereafter GL98; Suleymanova \& Pugachev 1998, 2002) accompanied
by prominent outer edge ``90$\deg$ flips'' in the position angle (PA).  
Earlier average studies (Manchester 1971; Morris \etal\ 1981; Bartel 
\etal\ 1982) together with the more recent single-pulse analyses of 
Gil \& Lyne (1995), von Hoensbroech \& Xilouris \etal\ (1997), and 
Mitra (1999) provide an unusually comprehensive picture of this 
pulsar's outer edge depolarization.  The phenomenon persists to the 
highest frequencies, as can clearly be seen in the 10.55-GHz profile 
of von Hoensbroech \& Xilouris above.

\begin{figure}
\begin{center}
\epsfig{file=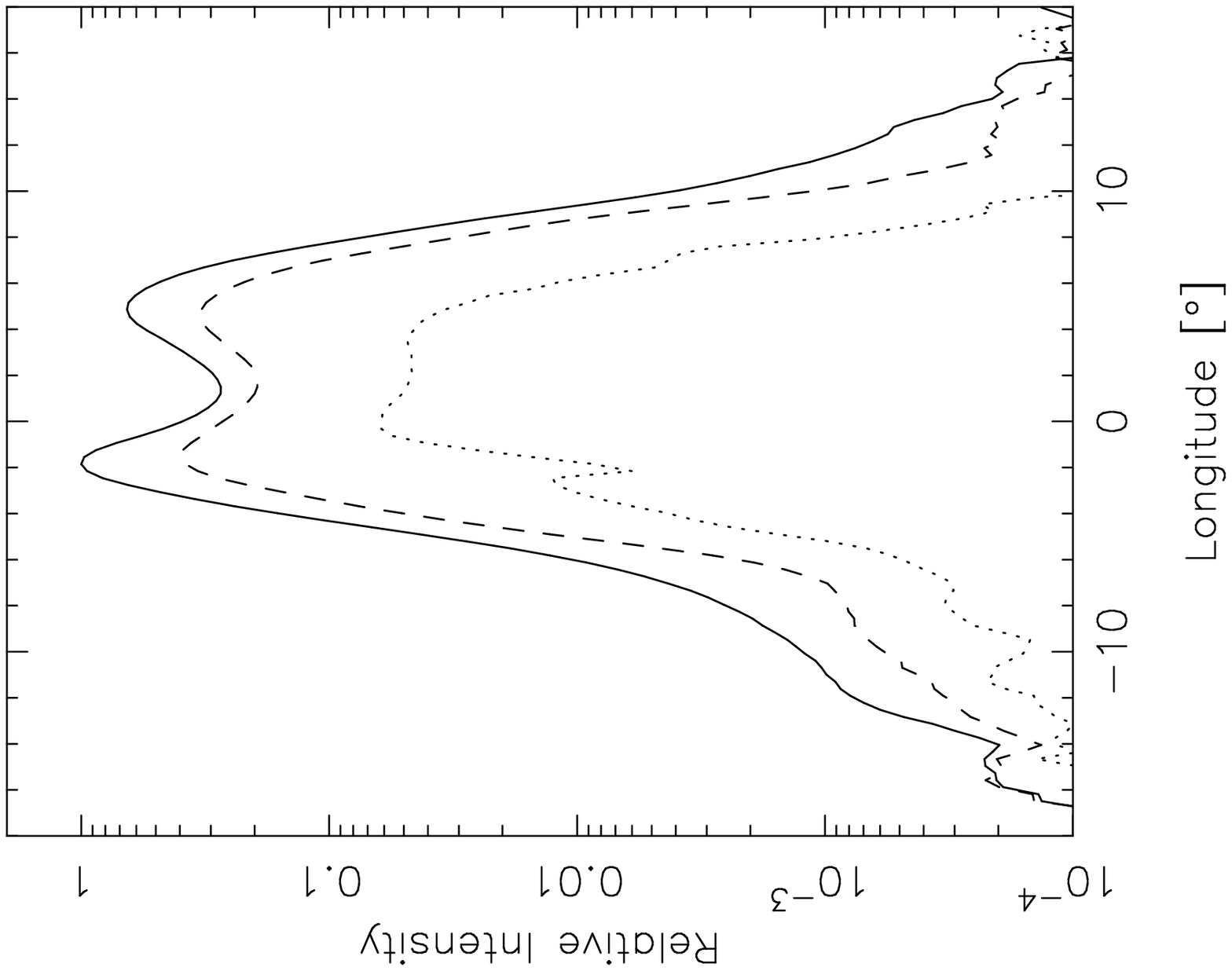,height=6.0cm,angle=-90}
\epsfig{file=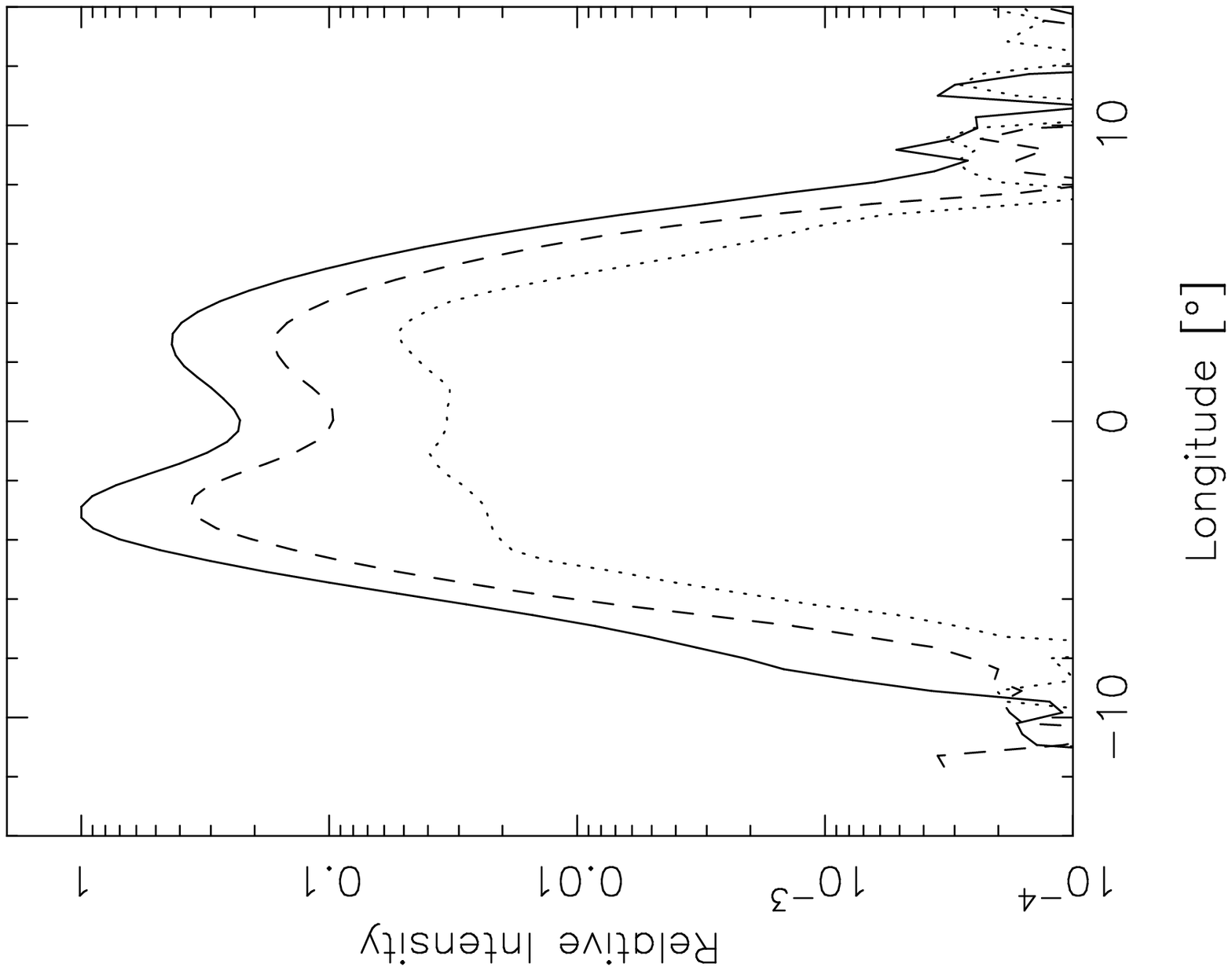,height=6.0cm,angle=-90}
\end{center}
\caption[]{Log-linear plots of the polarized intensity
profiles of pulsar B1133+16, showing the full extent of
the edge depolarization---430 MHz (upper) and 1414 MHz
(lower).  The total intensity, Stokes parameter $I$,
is given by the solid curve, the linear polarization
$L$ by the dashed curve, and $-V$ by the dotted one
(only rh circular is observed in this pulsar).  The
three respective curves were smoothed over five samples,
normalized to the maximum in $I$, and the statistical
bias in $L$ was removed.  The 430-MHz and 21-cm sequences
had lengths of 956 and 2180 pulses, respectively.}
\label{fig1}
\end{figure}

Other obvious exemplars are pulsars B0525+21 and 1133+16, which
clearly exhibit the outer edge depolarization phenomenon over the
entire range of frequencies that they can be observed [see the above
papers as well as Blaskiewicz \etal\ (1991), von Hoensbroech (1999),
and Weisberg \etal\ (1999)].  For 0525+21, which has a more central
sightline traverse (see Table 1), individual-pulse polarization
displays show that the weaker secondary polarization mode dominates
the primary one (hereafter SPM and PPM, respectively) only on the
extreme outer edges of its profiles; whereas for 1133+16, which has a
more oblique sightline traverse, SPM dominated samples can be seen
over a larger longitude range.\footnote{In this paper, the terms 
primary (PPM) and secondary (SPM) polarization mode denote little 
more than their relative strength.}

Reference to the now extensive body of published average polarimetry
provides several hundred examples of pulsars whose conal component
pairs have prominently depolarized outer edges.  The effect is so
widespread, indeed, that it is difficult to identify completely
convincing examples to the contrary.  The stars comprising four of the
five main profile classes (\eg, see Paper VI) of conal single ({\bf
S$_{\rm d}$}), double ({\bf D}), triple ({\bf T}), and five-component
({\bf M}) virtually all exhibit the phenomenon as do the few stars in
the more restricted {c\bf T} and {c\bf Q} classes.  It is worth noting
that good examples of outer edge depolarization are found among stars
with both inner and outer conal configurations; all three of the stars
with inner cones discussed in the foregoing Paper VII [Mitra \& Rankin
(2002)] show the effect, though interestingly, in each case it is 
more prominent on the trailing than on the leading edge.\footnote{As 
discussed in Paper VII, many or most stars with inner-cone profile
configurations also exhibit discernible emission in the ``baseline''
region, far in advance of the leading component and sometimes after the 
trailing component, perhaps because a weak outer cone is also emitted.} 
The best examples of stars with little or no edge depolarization all
either have (or probably have, given that some are yet not well 
observed) core single {\bf S$_{\rm t}$} or inner-cone triple {\bf T}
configurations; some are B0355+54, 0450+55, 0540+23, 0559--05,
0626+24, 0740--28, 0833--45, 0906--49 (main pulse), 1055--52, 1322+83,
and 1737--30, and note that, overall, these stars have much shorter
periods than is typical for the normal pulsar population.

We then summarize the overall characteristics of the edge-depolarization 
phenomenon: 
\begin{enumerate}
\item The average linear polarization $L$ (=$\sqrt{Q^2+U^2}$) falls
off much faster than the total power $I$ on the edges of the profile
and decreases asymptotically to near zero.
\item This phenomenon usually
occurs over a very broad band---essentially the entire range of the
observations---and therefore, the edge depolarization appears to be
nearly independent of frequency.
\item This profile-edge depolarization
is modal in origin, meaning that it largely occurs through the
incoherent addition of PPM and SPM power both within samples and from
pulse to pulse.  
\item The edge depolarization affects conal component
pairs, and therefore must be regarded as a roughly symmetrical,
structural feature of conal emission beams.  
\item Then, in terms of such beams: 
    \begin{itemize} 
      \item the outer edge depolarization requires that the modal power
             be about equal at large angles to the magnetic axis, and 
      \item the proximity of the depolarized outer edges of conal 
            component pairs to their more highly polarized inner edges 
            requires that the weaker mode peak at slightly larger angles 
            to the magnetic axis than the stronger one.
    \end{itemize}
\end{enumerate}

In the remainder of this paper we will explore the causes and
consequences of these circumstances, drawing extensively on the
earlier articles of this series, Papers I--VII (see References).  We
will show that these structural characteristics of conal emission
beams---and therefore well resolved conal component pairs---are almost
certainly the result of subbeam circulation as in pulsar B0943+10 (see
Deshpande \& Rankin 2001).  This circulation, in sweeping a series of
polarized subbeams around the magnetic axis and past our sightline, is
responsible for the outer-edge depolarization and (sometimes periodic)
modal fluctuations in pulsars where the sightline traverse cuts the
emission beam centrally ({\it e.g.}, 0525+21); it is also largely
responsible for the very different polarization effects observed in
pulsars where the sightline traverse is oblique ({\it i.e.}, B0809+74).  
Of course, only subbeams with particular angular polarization patterns
can produce the particular sorts of depolarized profile forms that are
observed, and in the remainder of this paper we endeavor to understand
what general features are required of them.  In the following sections, 
we first briefly describe our observations and then consider the
contrasting characteristics of stars, first with well separated conal
component pairs, and then with conal single {\bf S$_{\rm d}$} profiles.  
The penultimate section gives the results of modelling the polarized
emission beam, and we conclude with a summary of our results and a
discussion of their implications.

\section*{Observations}

\begin{deluxetable}{ccccccc}
\small
\tablewidth{0pt}
\tablecaption{Pulsar Parameters.\label{tbl-1}}
\tablehead{Pulsar&$P$&$\beta/\rho$\tablenotemark{a}&$f$&Source&Date&BW \\
  (B--)    &  (s)  &        & (MHz) &      &  & (MHz)      }
\startdata
 0301+19   & 1.388 &  0.45  &  430 &  AO   &1974Jan5  & 10/32\\
 0329+54   & 0.715 &  0.31  &  840 & WSRT  &2002Jan10 & 80/512\\
 0525+21   & 3.745 &  0.19  &  430 &  AO   &1974Apr4  & 10/32\\
 1133+16   & 1.188 &  0.78  &  430 &  AO   &1992Oct19 & 10/32\\
           &       &        & 1414 &  AO   &1992Oct15 & 20/32\\
 1237+25   & 1.382 &$\sim$0 &  430 &  AO   &1974Jan6  & 10/32\\
 2020+28   & 0.343 &  0.49  &  430 &  AO   &1992Oct16 & 10/32\\[4pt]
%
 0809+74   & 1.292 &$ 0.93 $&  328 & WSRT  &2000Nov26 & 10/64\\
 0820+02   & 0.865 &$ 0.98 $&  430 &  AO   &1992Oct19 & 10/32\\
 0943+10   & 1.098 & -1.01  &  430 &  AO   &1992Oct19 & 10/32\\
 1923+04   & 1.074 &$ 0.97 $&  430 &  AO   &1991Jan6  & 10/32\\
 2016+28   & 0.558 &$ 0.96 $&  430 &  AO   &1992Oct15 & 10/32\\
 2303+30   & 1.576 &$ 0.99 $&  430 &  AO   &1992Oct15 & 10/32\\
\enddata
\tablenotetext{a}{The sign of the magnetic impact angle $\beta$ is 
specified only when it is known.  $\rho$, the conal beam radius, is 
positive definite.  The values refer to 1 GHz and most are taken 
from Rankin (1993a,b; hereafter, Paper VI).}
\end{deluxetable}

The source and character of our observations are summarized in 
Table 1.  The Arecibo Observatory (AO) recordings were made 
under two polarimetry programs, the first in the early 1970s 
and the second in 1992, and both are described in Rankin \& 
Rathnasree (1997).  The 328- and 840-MHz sequences were made 
using the Westerbork Synthesis Radio Telescope (WSRT) with 
its pulsar machine {\tt PuMa}, and these are described in 
Ramachandran \etal\ (2002).

\section*{The Depolarization Pattern of Conal Component Pairs}
\label{sec-depolar}

\begin{figure*}
\begin{center}
\begin{tabular}{@{}lr@{}lr@{}}
{\mbox{\epsfig{file=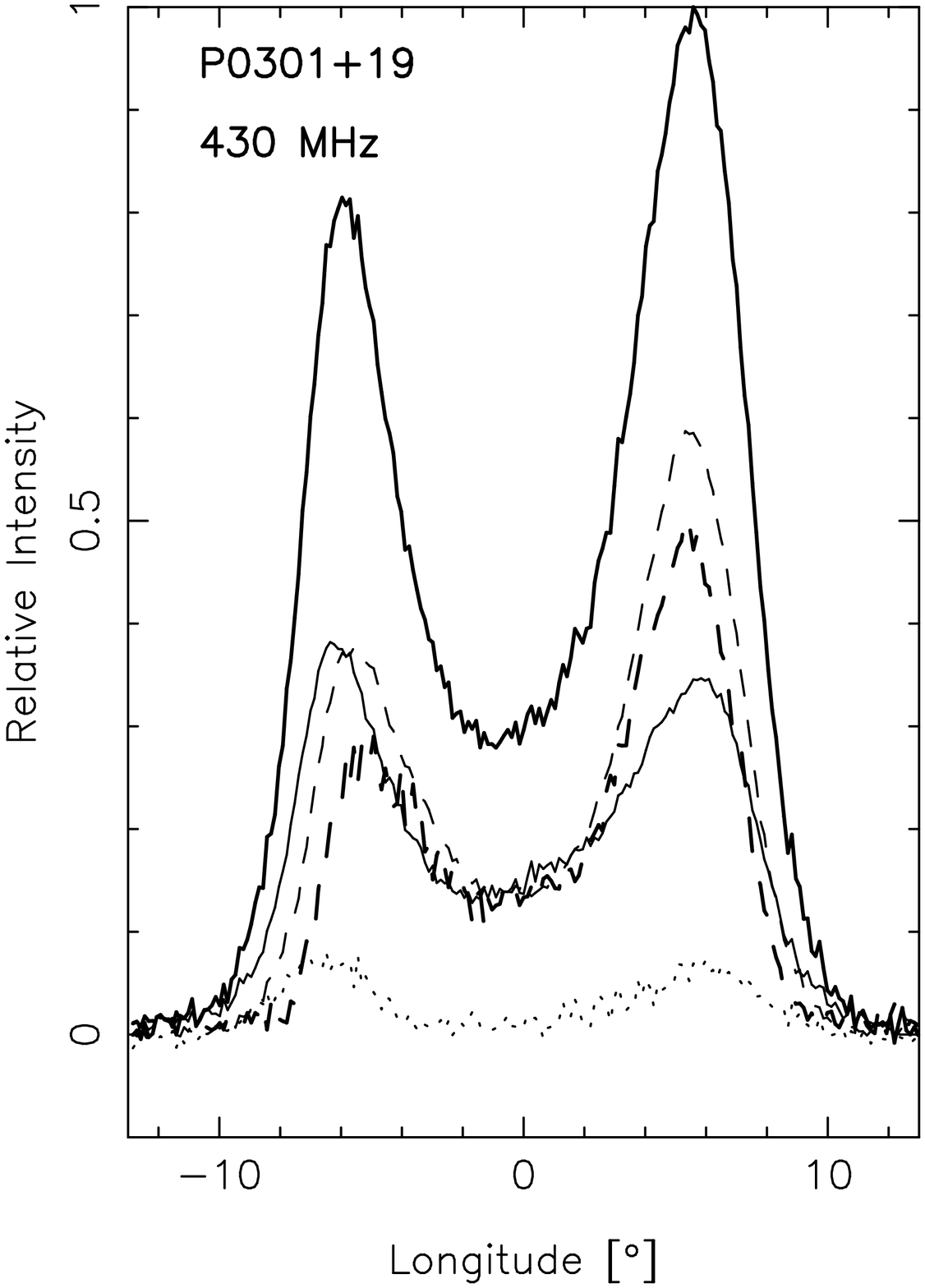,height=7.3cm}}}&
{\mbox{\epsfig{file=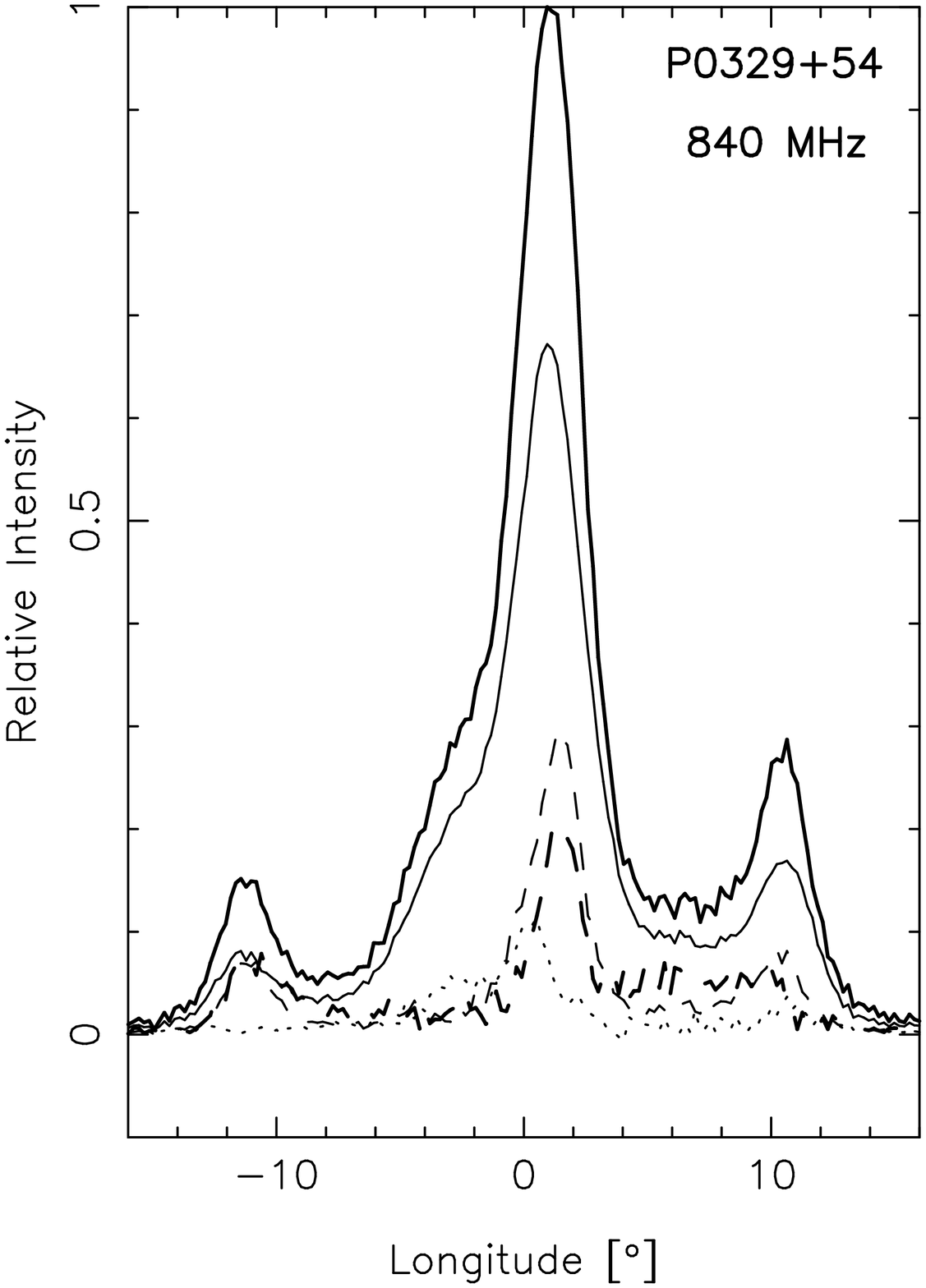,height=7.3cm}}}&
{\mbox{\epsfig{file=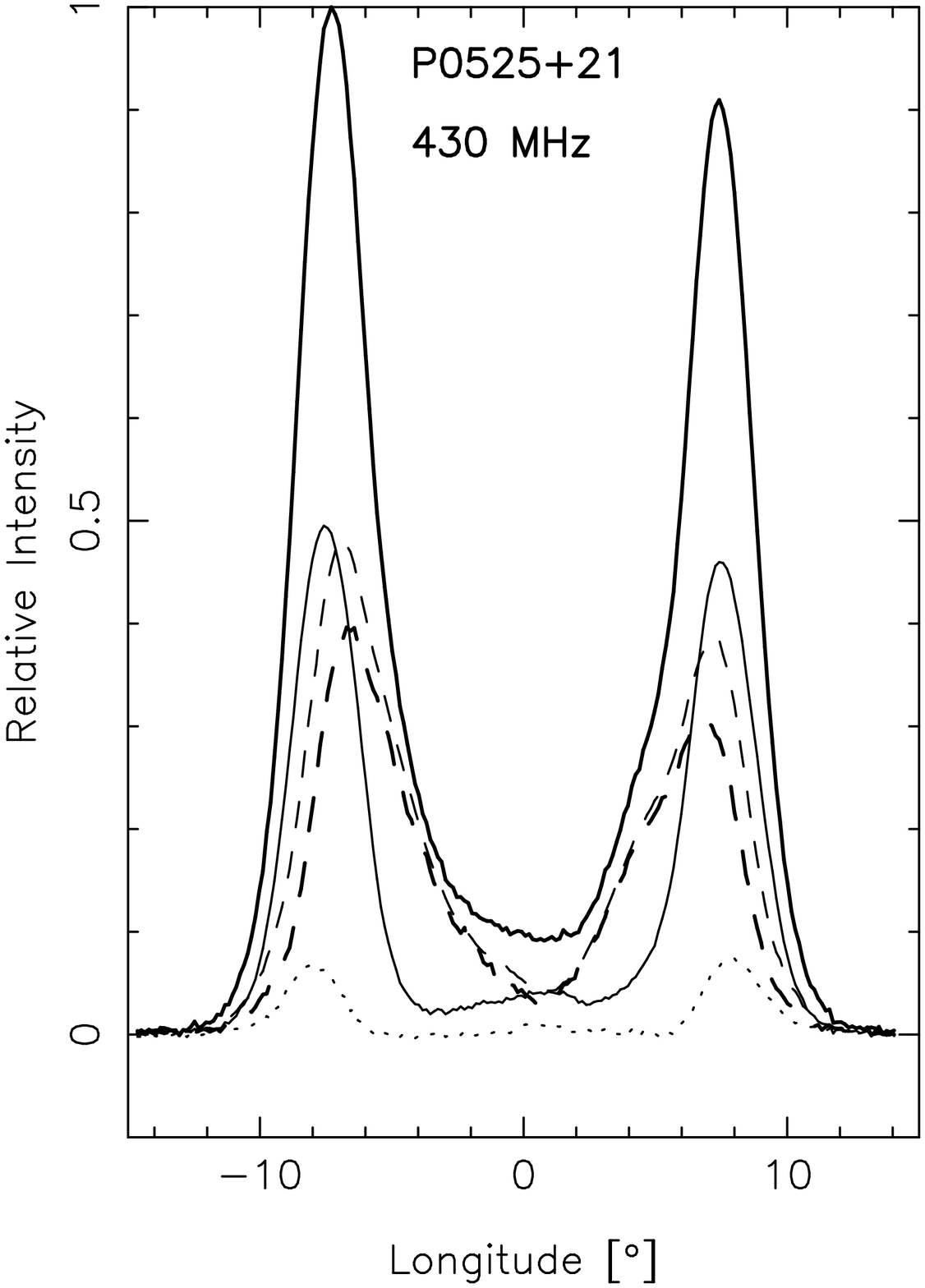,height=7.3cm}}}\\
{\mbox{\epsfig{file=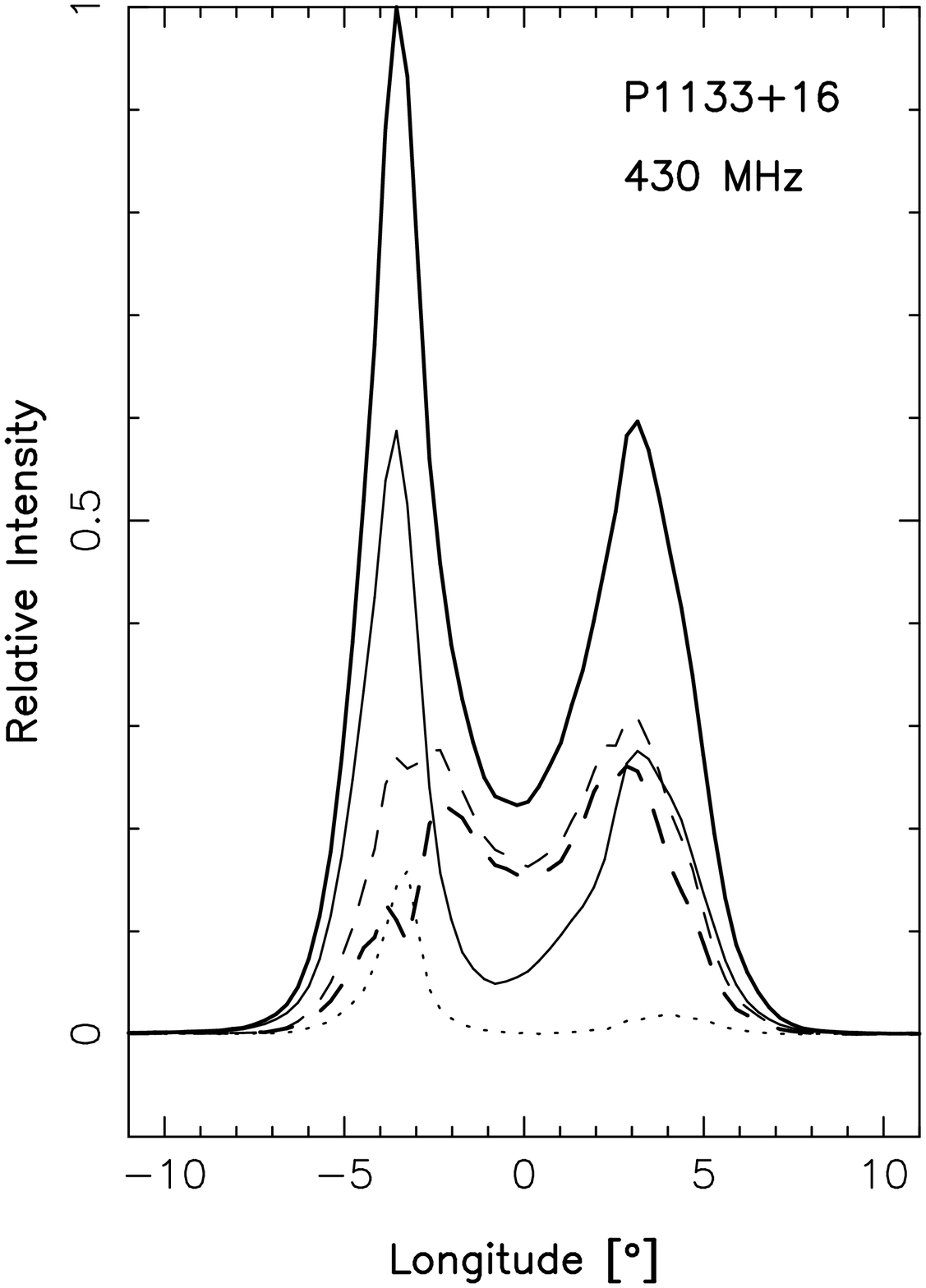,height=7.3cm}}}&
{\mbox{\epsfig{file=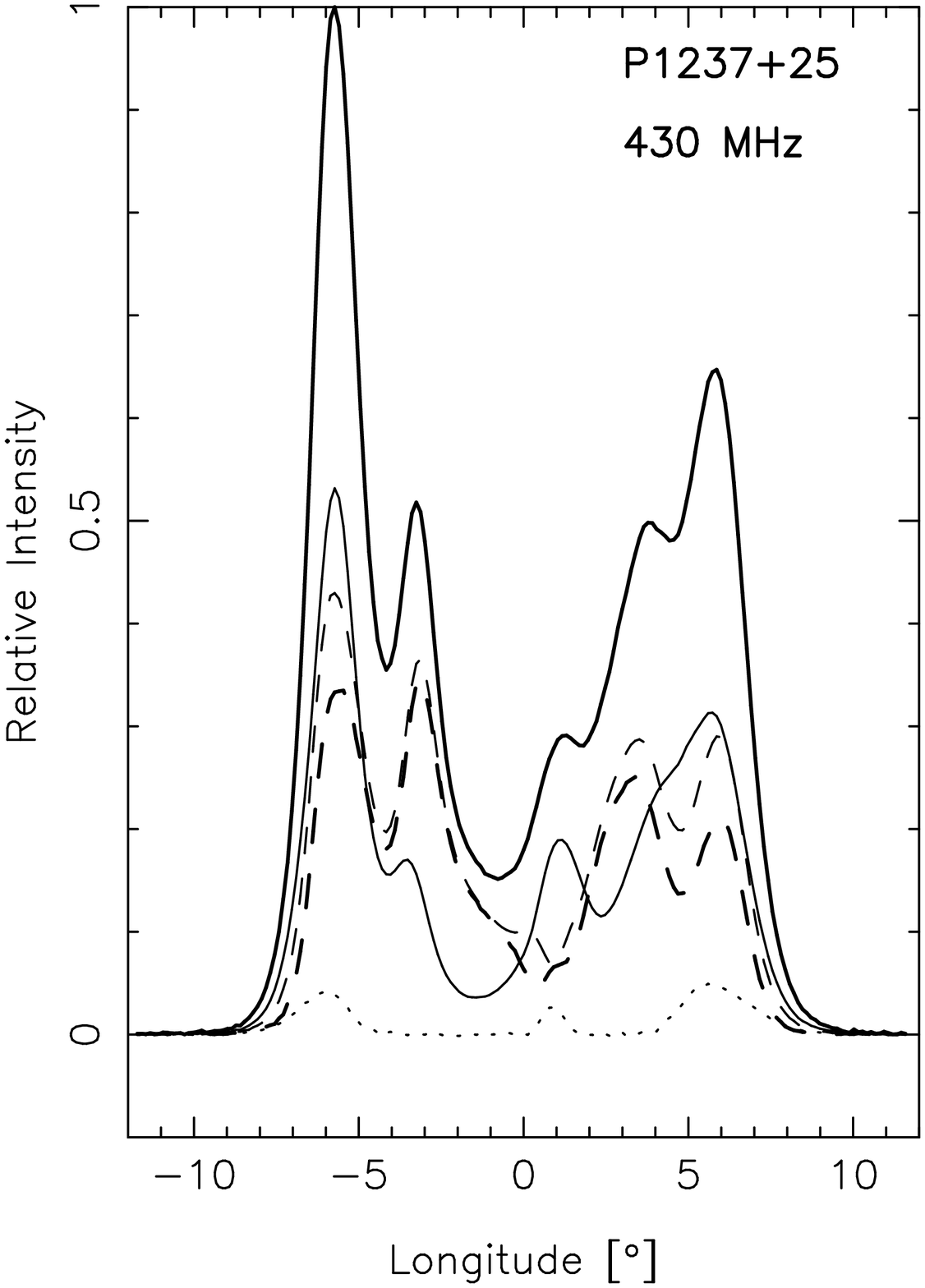,height=7.3cm}}}&
{\mbox{\epsfig{file=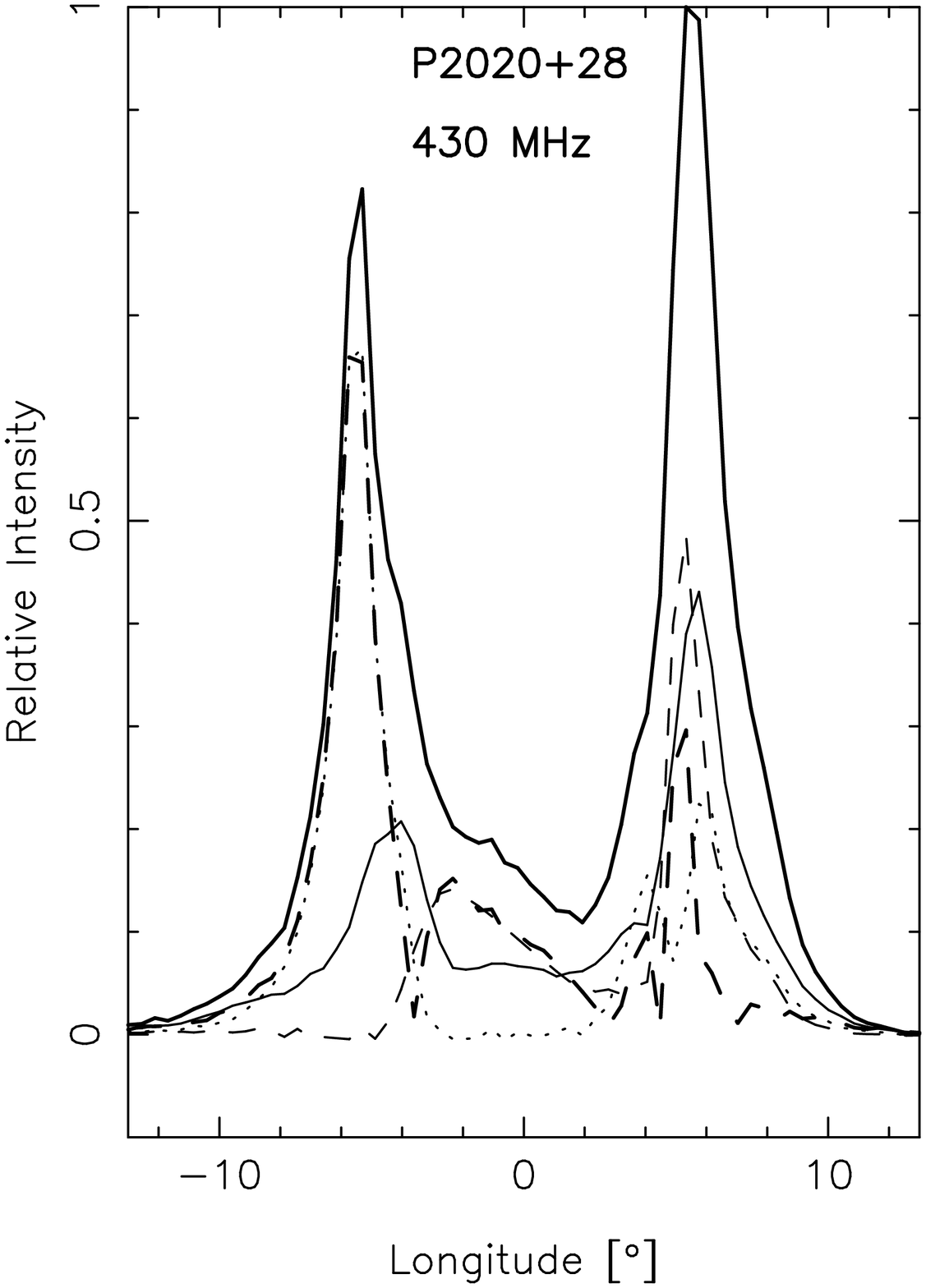,height=7.3cm}}}\\
\end{tabular}
\end{center}
\caption[]{Three-way, mode-segregated average profiles
for pulsars with prominent conal component pairs, B0301+19,
0329+54, 0525+21, 1133+16, 1237+25, and 2020+28.  The
heavier solid and dashed curves give the total power
(Stokes $I$) and total linear $L$; whereas, the lighter
dashed, dotted and solid curves give the PPM, SPM and UP
power, computed according to the algorithm in Deshpande \&
Rankin's (2001) Appendix (see text).}
\label{fig2}
\end{figure*}

\begin{figure*}
\begin{center}
\begin{tabular}{@{}lr@{}lr@{}}
{\mbox{\epsfig{file=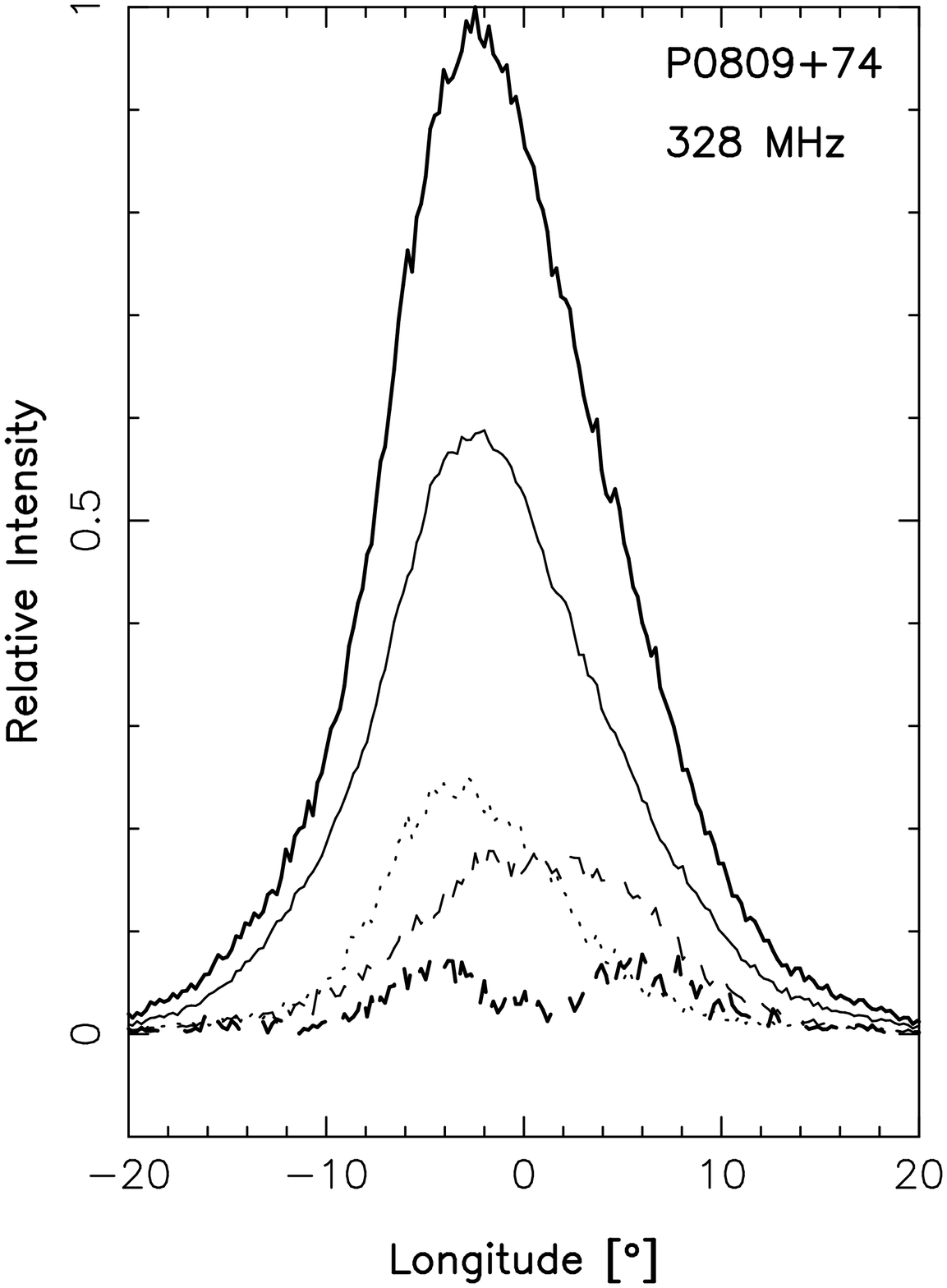,height=7.3cm}}}&
{\mbox{\epsfig{file=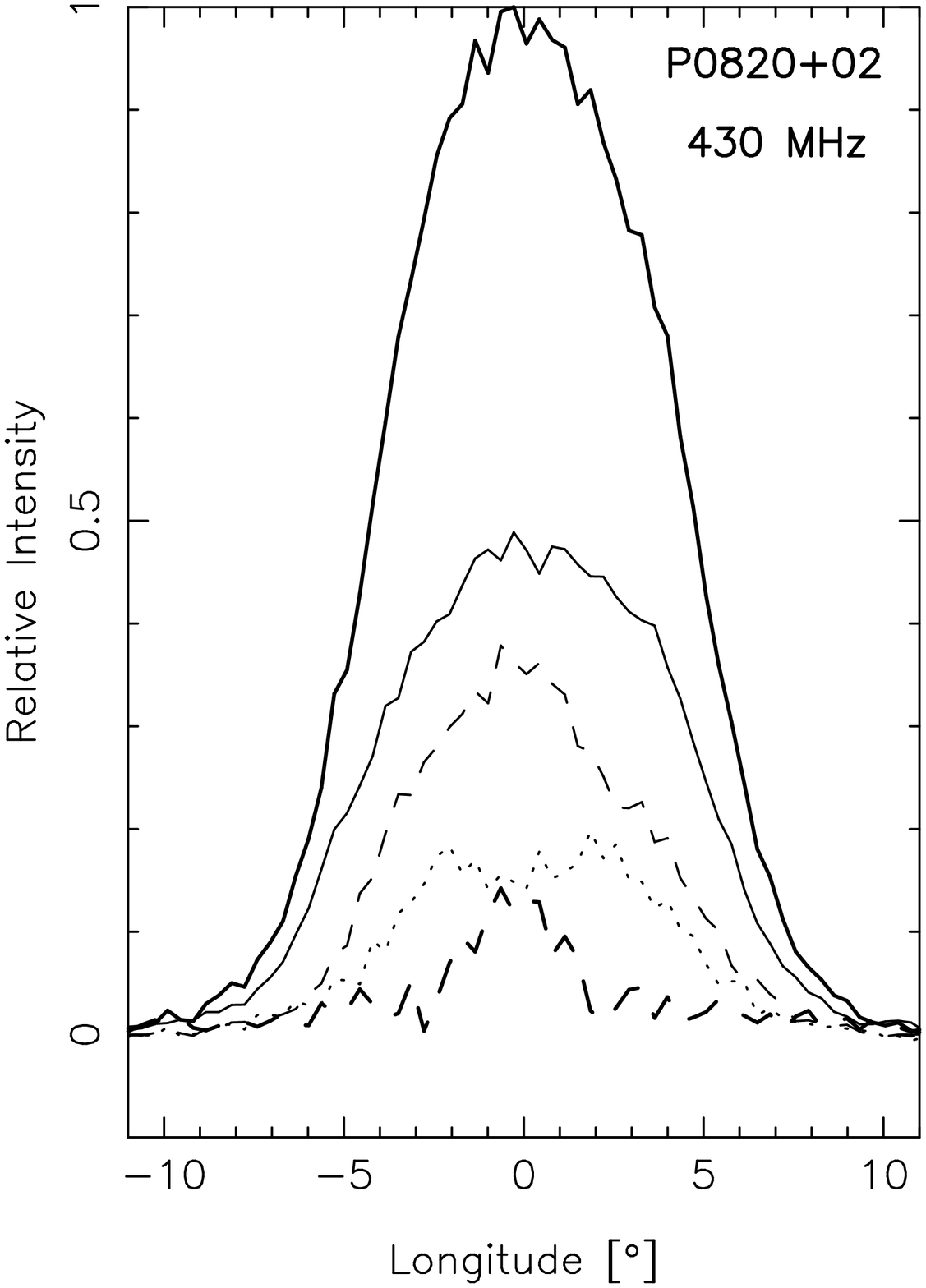,height=7.3cm}}}&
{\mbox{\epsfig{file=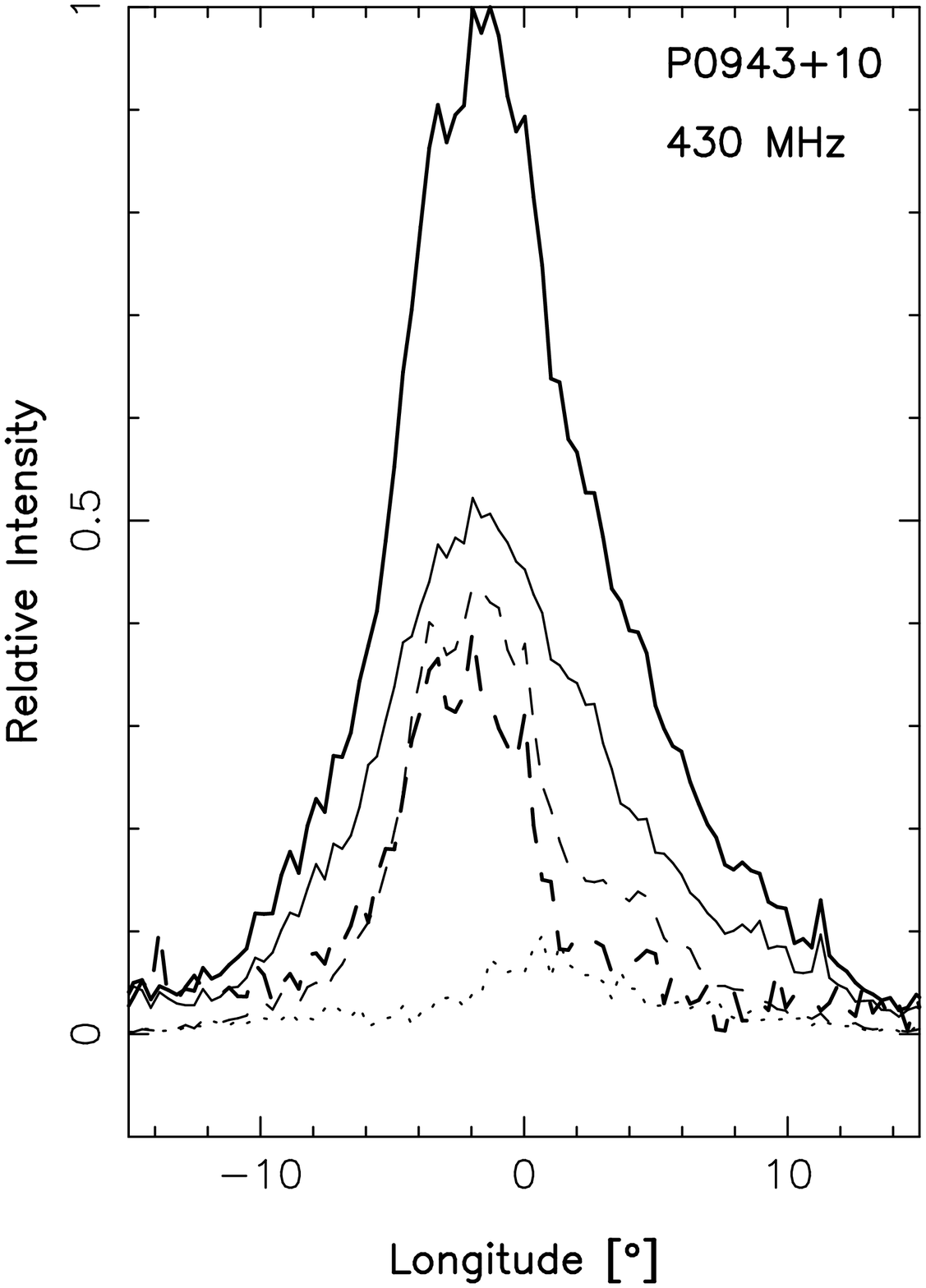,height=7.3cm}}}\\
{\mbox{\epsfig{file=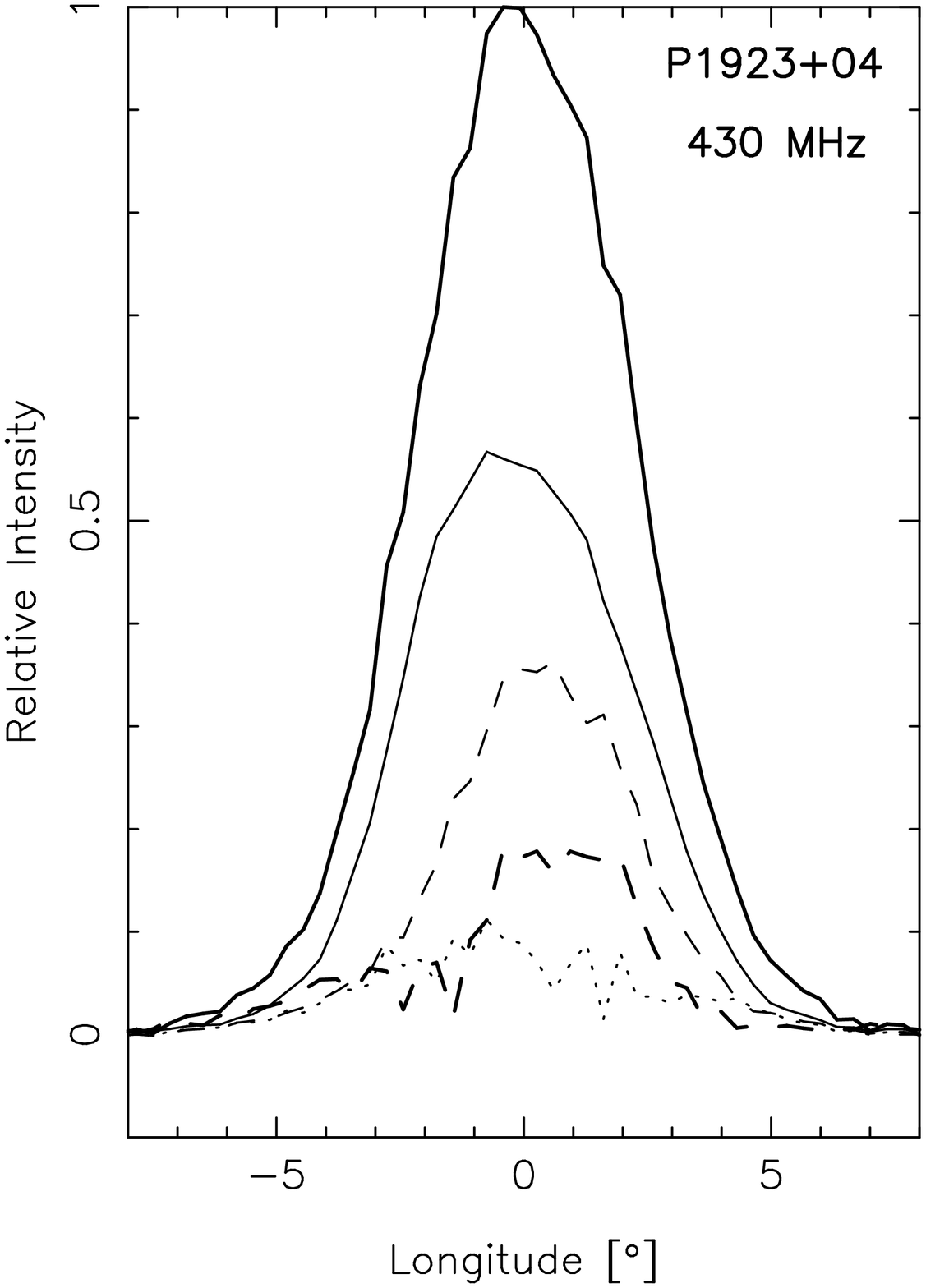,height=7.3cm}}}&
{\mbox{\epsfig{file=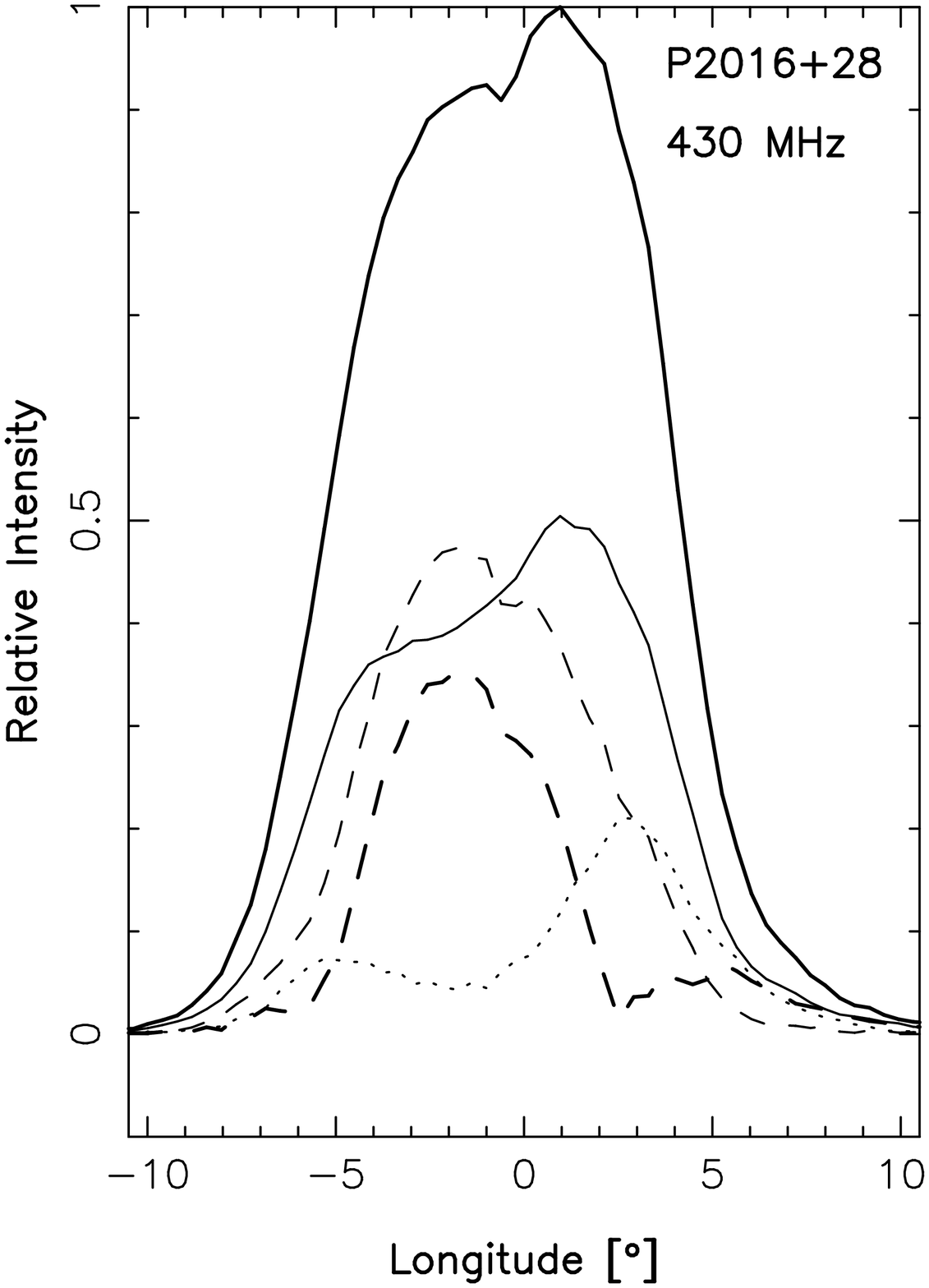,height=7.3cm}}}&
{\mbox{\epsfig{file=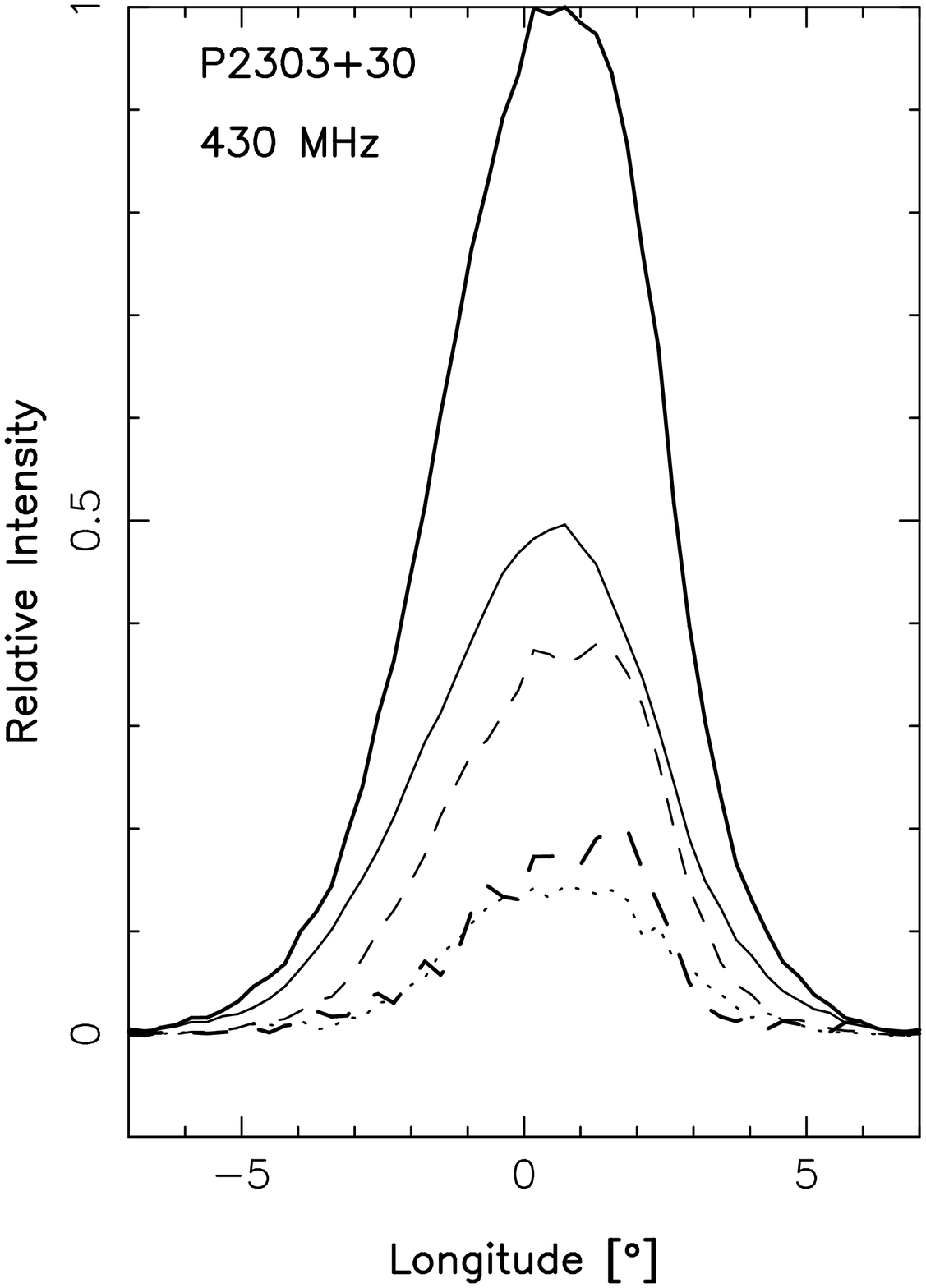,height=7.3cm}}}\\
\end{tabular}
\end{center}
\caption[]{Three-way, mode-segregated average profiles for
six conal single {\bf S$_{\rm d}$} pulsars, B0809+74, 0820+02,
0943+10, 1923+04, 2016+28, and 2303+30 as in Fig.~\ref{fig2}.}
\label{fig3}
\end{figure*}

\begin{figure*}
\begin{center}
\epsfig{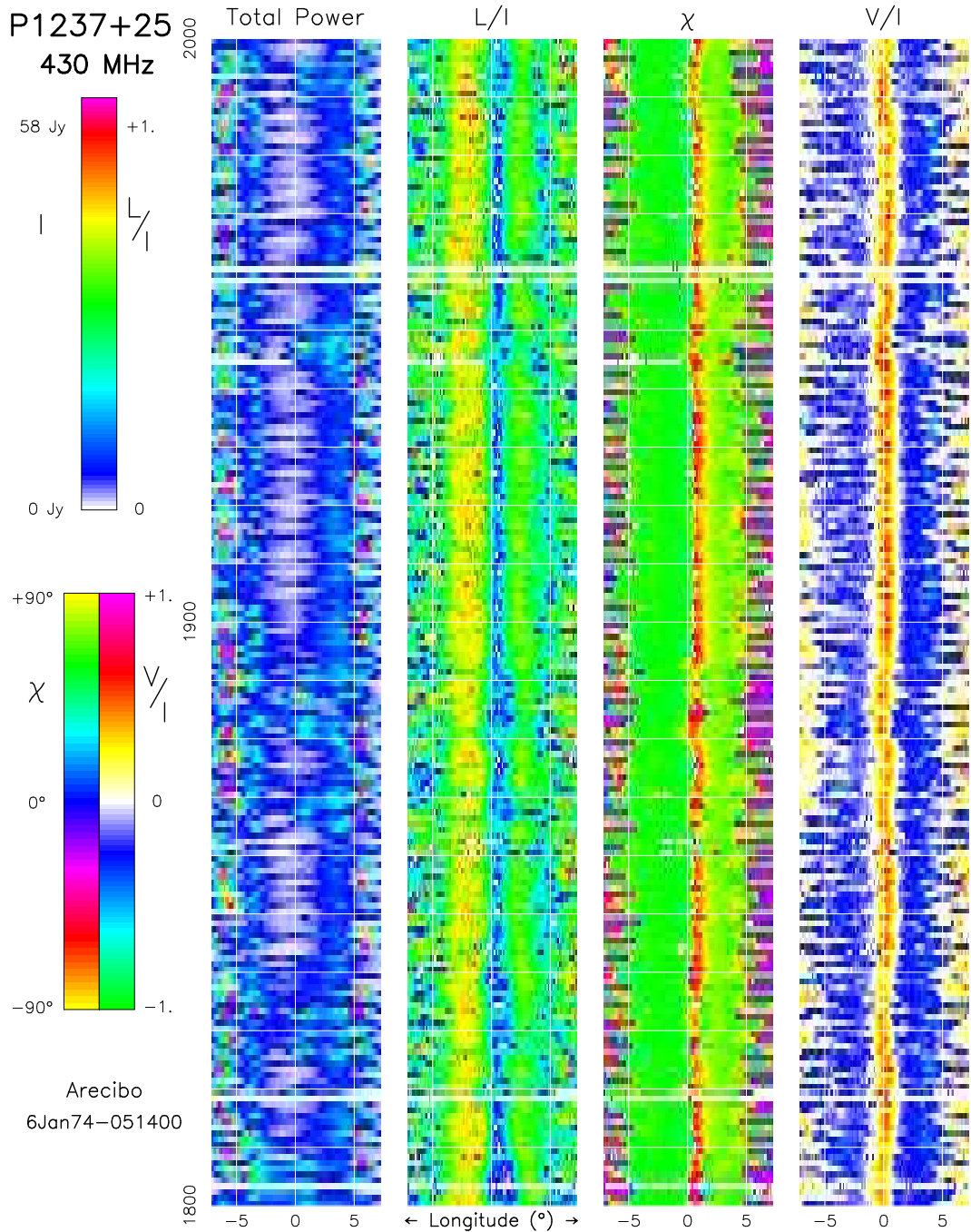}
\end{center}
\caption[]{Color polarization display of a 200-pulse portion of the
430-MHz observation in Figs.  2 and 5.  The first column gives the
total intensity (Stokes $I$), with the vertical axis representing the
pulse number and the horizontal axis pulse longitude, colour-coded
according to the left-hand scale of the top bar to the left of the
displays.  The second and third columns give the corresponding
fractional linear polarization ($L/I$) and its angle ($\chi={1\over
2}\tan^{-1}{U/Q}$), according to the top-right and bottom-left scales. 
The last column gives the fractional circular polarization ($V/I$),
according to the bottom-right scale.  Plotted values have met a
threshold corresponding to 2 standard deviations of the off-pulse
noise level.  Note the 2.63 c/$P_1$ modulation associated with the
outer conal component pair---and that this modulation has a strikingly
modal character as can be seen particularly clearly in the orthogonal
chartreuse and magenta PAs.}
\label{colourplot}
\end{figure*}

Let us now look in more detail at the manner in which the outer edges
of conal component pairs are depolarized.  Turning first to pulsar
1133+16, Figure~\ref{fig1} shows the relative behaviors of the
logarithms of Stokes parameters $I$, $L$, and $-V$ as a function of
longitude for a 430-MHz sequence (top) and a 21-cm sequence (bottom). 
Here we can follow the behavior of the fractional polarization far out
into the ``wings'' of the star's profile.  We see not only that the
depolarization persists to very low intensity levels, but also that
its linear and circular polarization generally decrease with or faster
than the total power down to the point where the noise fluctuations
begin to dominate at 2--4$\times$10$^{-4}$ (note that only the
absolute value of the noisy quantities can be plotted).

Probably, this behavior is typical of many pulsars, but only for a
few, such as 1133+16, can polarized profiles with such a large dynamic
range be computed.  Even for 1133+16 it would be interesting to
compute a more sensitive such display.  These observations from
Arecibo were only some 40 minutes long, so with care it should be
possible to reduce the relative noise level much further.  If, then,
it is generally true that the outer edges of conal profiles---and thus
the outer edges of conal beams---are accurately depolarized on average, 
it provides a strong constraint on the angular beaming characteristics 
of the modal emission.

We can look at this outer-edge depolarization in more detail by
conducting an appropriate mode-segregation analysis on selected
sequences.  Two such algorithms were described in Deshpande \&
Rankin's (2001) Appendix, and we use here the three-way mode
segregation method, because it provides the greatest flexibility.  
It produces two fully polarized PPM and SPM pulse sequences and 
a fully depolarized UP sequence, while making no restrictive 
assumptions about the origin of the depolarization.  Briefly, 
the $I$ and $L$ of each sample are compared with a noise threshold, 
and its respective $L$ and $I$--$L$ portions accumulated in three 
partial sequences depending on whether the sample is PPM dominated, 
SPM dominated (relative to a model PA traverse which defines the 
former), or essentially unpolarized (UP).

The results of these analyses for pulsars with prominent conal
component pairs are given in Figure~\ref{fig2}, where the heavier
curves give the usual total-power ($I$) and total linear ($L$)
profiles, while the lighter curves show the PPM (dashed curve), SPM
(dotted curve) and UP (solid curve) profiles.  A similar (but more
primitive) analysis for B1737+13 (Rankin \etal\ 1986) can also be
compared, as can the excellent modal polarization studies of B2020+28
and 0525+21 by McKinnon \& Stinebring (1998, and 2000, respectively;
hereafter MS98 and MS00).  The sources of the various sequences are
given in Table~\ref{tbl-1}, where we also tabulate $\beta/\rho$.  For
the stars considered in this section, $|\beta|/\rho < 0.8$, a
sightline geometry which produces well resolved conal component
pairs.  Note, by contrast, that the conal single {\bf S$_{\rm d}$}
pulsars considered in the next section all have $|\beta|/\rho > 0.9$.

For most of the stars (all but B2020+28), we see a fairly consistent
picture. The weaker mode only rarely has sufficient intensity to
dominate a sample, so the aggregate SPM power is typically only some
10\% that of the PPM---and almost all of this SPM power is found on
the ``wings'' of the profiles.  Often, the SPM power peaks slightly
further out than the PPM and exhibits a narrower angular width.  Note,
further, that the UP distribution behaves very similarly to both the
PPM and SPM curves, so we may view some portion of the UP power as the
accumulation of samples which were depolarized by equal contributions
of PPM and SPM power---and indeed, the UP curves always asymptotically
approach the overall $I$ curves at very low power levels.  This
behavior could also be demonstrated by applying the two-way modal
``repolarization'' technique in Deshpande \& Rankin, which proceeds
under the assumption that the depolarized samples contain equal PPM
and SPM levels of power.

Though pulsar 2020+28's modal behavior appears more complex ({\it e.g.}, 
Cordes \etal\ 1978; MS98), we see many of the same features---for
instance, that the UP power approaches the total power on the extreme
edges of profile.  Indeed, MS98's analysis based on ``superposed
modes'' suggests similar conclusions.  The well measured profile
demonstrates that its ``two'' components each have a good deal of
structure---seen as ``breaks'' in the total-power curves---but the
PPM, SPM and UP curves demonstrate, in addition, that much of the
complexity is modal in origin.  The complex modal behavior of this
pulsar deserves much fuller study, and a well measured polarimetric
pulse sequence in the 100--200-MHz range would add much to our
knowledge.

Overall, we see that the conal component pairs depicted in
Figure~\ref{fig2} all have moderate to high levels of fractional
linear polarization---that is, typically some 50\%---though most have
narrow, interior regions of longitude where the linear polarization is
higher.  We shall see that this stands in sharp contrast to the {\bf
S$_{\rm d}$} pulsars considered in the following section.  Our point
is that when $|\beta|/\rho$ is relatively small---producing well
resolved conal component pairs---the mode mixing depolarizes the outer
edges, but not the profile interior.  This then reflects
properties---both dynamic in terms of modulation phenomena and
polarizational---of the conal emission beam, and we must reflect on
just how this is possible.

\section*{The Depolarization Patterns of Conal Single Stars}

We have just considered a group of stars in which our sightline makes
a fairly central traverse through their emission cone(s), and we now
turn to members of the conal single {\bf S$_{\rm d}$} group, all of
which are configured by a tangential traverse along the average
emission cone.  Here we have the opportunity both to explore the conal
depolarization phenomena in a very different geometrical context and
then to investigate how the modulation and depolarization phenomena
are connected.  Figure~\ref{fig3} gives mode-segregated polarization
plots (similar to those in Fig.~\ref{fig2}) for six {\bf S$_{\rm d}$}
stars.  Here, it is important to keep in mind that each of these
pulsars has prominent ``drifting'' subpulses, so that the profiles
give only a static average of the subpulse polarization.  The displays
of Fig.~\ref{fig3} show that the UP (perhaps, mode-mixed) power is
typically 50\% of the total, so that the overall modal contributions
are comparable and the aggregate linear polarization is often small. 
While all of the total-power profiles are roughly unimodal (only
B0820+02 is really symmetrical), the modal profiles are more complex;
the two peak at different longitudes in B0809+74; the SPM has a double
form in B2016+28; and we have already noted the peculiar ``triple''
form of the aggregate linear in pulsar 0820+02.

As a class, the {\bf S$_{\rm d}$} stars exhibit conspicuously
depolarized profiles at metre wavelengths.  Indeed, this has been one
of the great obstacles to understanding their characteristics,
because, for many ({\it i.e.}, 0809+74), the modal complexity and low
fractional linear polarization make it difficult to accurately
determine even such a simple parameter as the PA sweep rate ({\it
e.g.}, Ramachandran \etal\ 2002).  Paradoxically, some also have
nearly complete linear polarization at certain longitudes and
frequencies ({\it i.e.}, as does 0809+74's leading edge at higher
frequencies) suggesting that mode-mixing is not always operative.

The {\bf S$_{\rm d}$} pulsars are also the profile class most closely
associated with the problematic phenomenon of ``absorption''.  It was
in 0809+74 that the effect was first identified (Bartel \etal\ 1981;
Bartel 1981)---that is, evidence that parts of the profiles were
``missing''---and strong evidence to this effect through subbeam-mapping 
methods have also been adduced for 0943+10 (Deshpande \& Rankin 2001). 
Surely one could imagine from 0943+10's asymmetric profile that a part
of its trailing-edge emission is ``absorbed'', though 0809+74's more
symmetric leading edge at meter wavelengths gives little clue that
emission appears to be missing here as well.  In short, the
circumstances defining the profile edges appear to be more complicated
for conal single stars than for the other species, and their modal
polarization characteristics are an aspect of this complexity.

\section*{What is the Relative Polarization-Modal Phase
in Conal Component Pairs?}
\label{sec-relative}

We have learned in the foregoing two sections that well resolved conal
component pairs are most depolarized on their extreme outer edges,
while the polarized modal emission in conal single stars accrues to
the depolarization essentially over the entire width of the profile. 
These circumstances begin to illuminate the polarization configuration
of the subbeams; and, indeed, we saw in Deshpande \& Rankin (2001;
fig.  19) that for 0943+10 the discernible SPM emission was found in
between the 20 PPM subbeams.  We have found a similar configuration
for pulsar 0809+74, where the PPM and SPM power centers are displaced
from each other systematically in both magnetic azimuth and colatitude
by perhaps 20\% of the subbeam spacing (Rankin \etal\ 2002).

A related question which has had no investigation at all is the
following: what is the modulation phase relationship between the PPM
and SPM power on the outside edges of pulsars with conal component
pairs?  Such a question is not trivial to answer because only a few of
such stars have modal modulation which is strongly periodic (while
virtually all of the {\bf S$_{\rm d}$} stars, for instance, in
Fig.~\ref{fig2} exhibit a good deal of regularity).  Two pulsars which
do have periodic modulation features are B1237+25 and 2020+28.

\begin{figure}
\begin{center}
\mbox{\epsfig{file=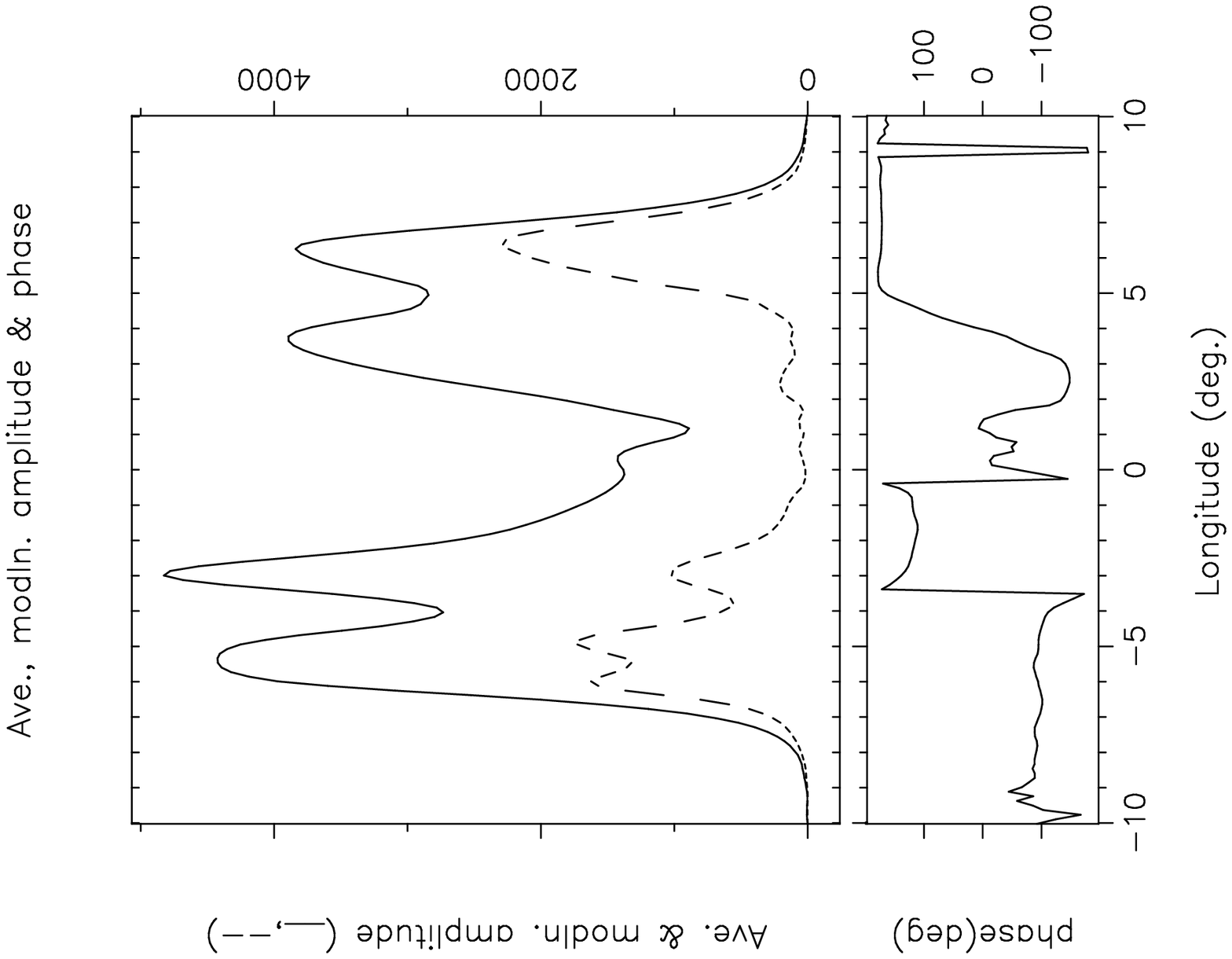,height=7.0cm,angle=-90.}}
\mbox{\epsfig{file=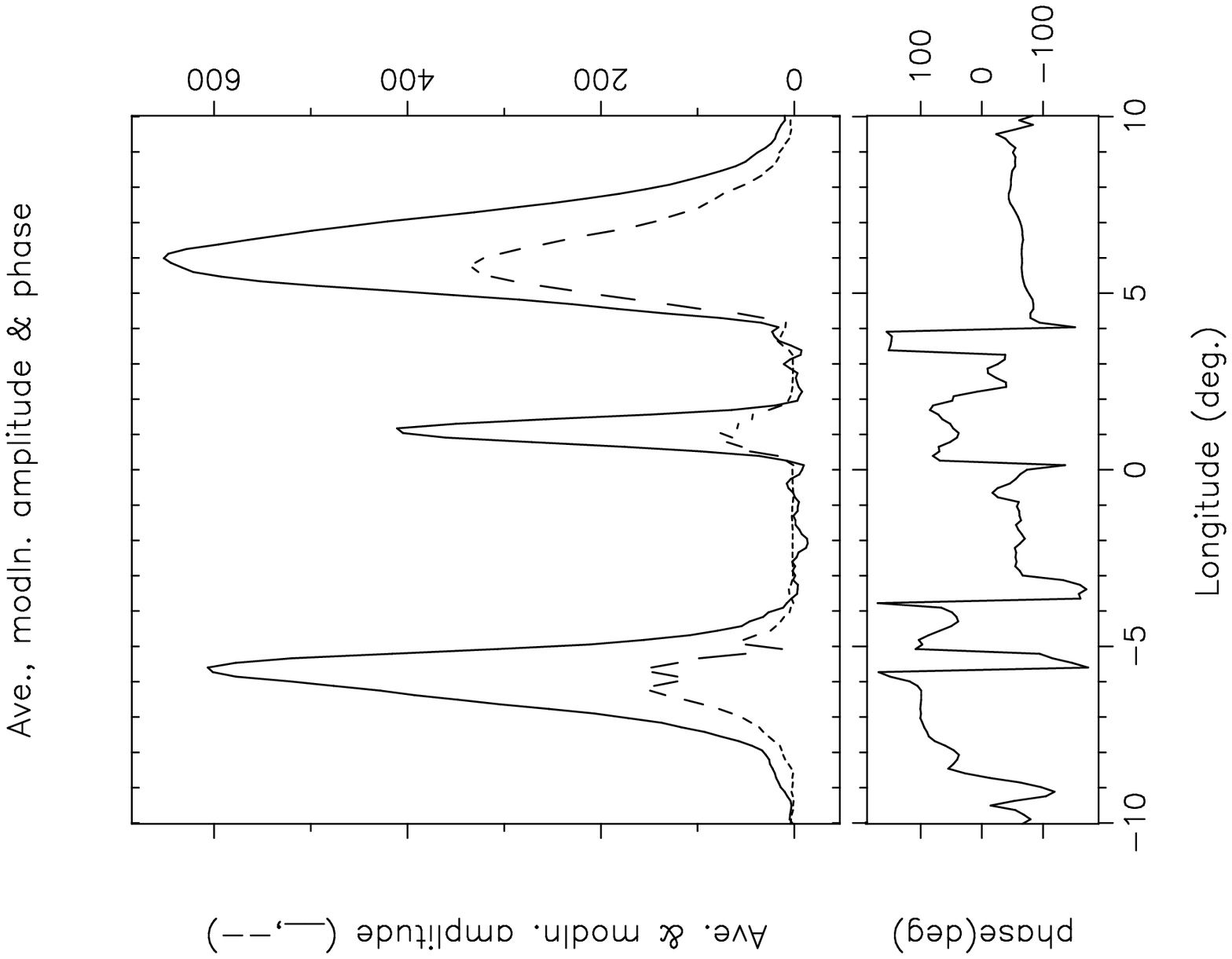,height=7.0cm,angle=-90.}}
\end{center}
\caption[]{Modulation amplitude and phase of the three-way segregated
PPM (top) and SPM (bottom) power in pulsar B1237+25 at 2.63 c/$P_1$. 
Note that 40--60\% of the fluctuation power under the outer cone is
modulated at this frequency and that the modal sequences have roughly
opposite phases.  The relationship at other longitudes is difficult to
interpret, because the mode segregation is less definitive and the
fluctuating power small or negligible.  The sequence here is a superset
of that in Fig.~\ref{fig2}.}
\label{fig4}
\end{figure}

Figure~\ref{colourplot} exhibits the character of this subpulse
modulation in pulsar B1237+25.  This sequence was chosen for its
brightness and relative freedom from nulls, and in consequence its
outer components show a particularly sharp feature at 2.63 c/$P_1$. 
This modulation can be seen very clearly in the first column which
gives the total power $I$.  The modal character of the modulation,
however, is most obvious in the third column depicting the PA, 
where the alternating magneta and chartreuse colors represent 
orthogonal PAs.  Further effects of this modal modulation can be 
seen in the varying levels of associated depolarization (second 
column) and the correlated variations of circular polarization 
(forth column).  Note that virtually all of these modal modulation 
effects are confined to the outermost pair of conal components, 
usually referred to as components I and V.

Figure~\ref{fig4} provides a more quantitative analysis of this modal
modulation under B1237+25's outer conal component pair.  The top
display gives the PPM profile after a three-way segregation of the
modal power along with a curve showing the fraction of this power
which is modulated at the feature frequency of 2.63 c/$P_1$; whereas
the lower panel gives the phase of this modulation.  As noted above,
we chose a part of the pulse sequence with few nulls, which also had a
particularly ``pure'' modulation feature.  Clearly, the phase is only
reliable under the outside conal component pair, where the modulation
represents a large fraction of the total modal power.  The lower
display gives similar information for the SPM-segregated partial
sequence.  Results for the UP partial sequence are irrelevant here and
thus not shown.

Remarkably, we see here that the PPM and SPM power are roughly out of
phase under the outer conal component pair.  The error in this phase
difference is relatively small as evidenced by the stable SPM phase
under the outer component pair.  Thus, when computed over the
256-pulse sequence, we have strong evidence that the modal power is
emitted in a manner which is far from ``in phase''.  This in turn
indicates that the modal power is systematically modulated, just as is
the total power.  Furthermore, that there is SPM power to segregate
implies (as can also be seen in Fig.~\ref{colourplot}) that, at times, 
the weaker SPM dominates the PPM.  

This behavior can be understood if both modes are, in general, present 
in every sample and combine incoherently---which is just the 
situation of ``superposed modes'' favored by MS00.

\section*{Geometry of Conal Beam Depolarization}

\begin{figure}[h]
\begin{center}
\mbox{\epsfig{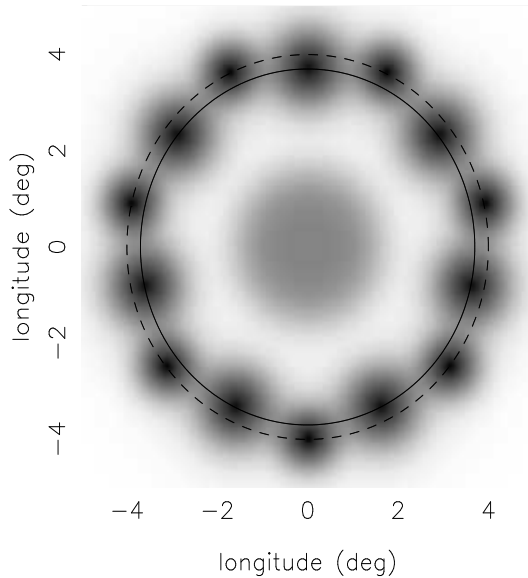}}
\end{center}

\caption[]{A gray scale representation of our rotating-subbeam 
model. Note the respective sets of PPM ($\parallel$ polarized) 
and SPM ($\perp$ polarized) ``beamlets'', which in turn comprise 
the PPM (inner, solid) and SPM (outer, dashed) subcones.  There 
is also an unpolarized, non-drifting central core component.}
\label{fig:simul_beam}
\end{figure}

\begin{figure}
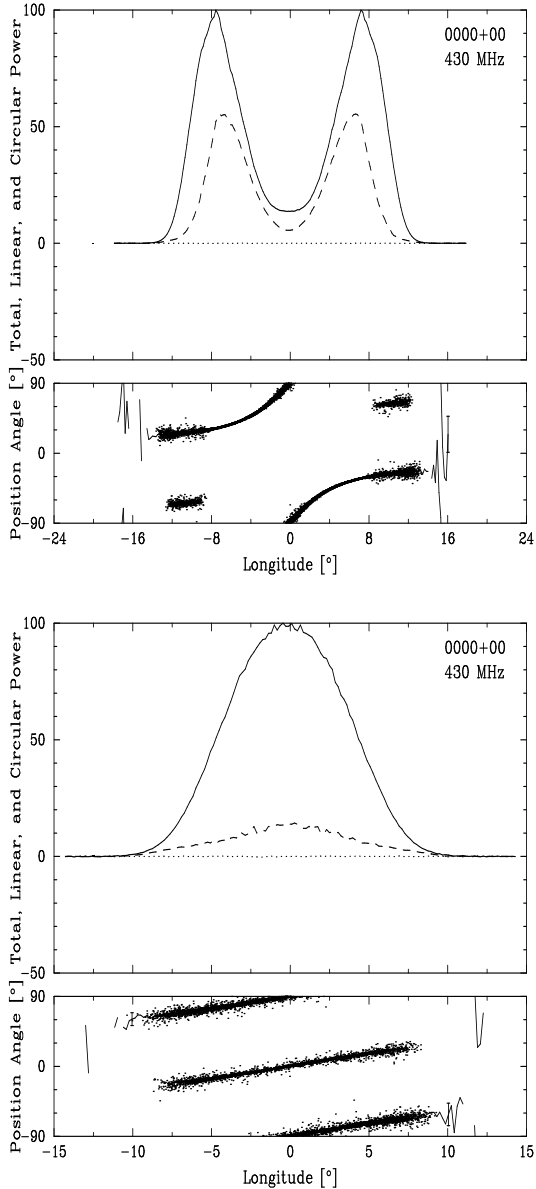

\begin{center}
\mbox{\epsfig{file=0525sim.ps,height=7.0cm,width=\linewidth,angle=-90}}

\vspace*{0.5cm}

\mbox{\epsfig{file=0820sim.ps,height=7.0cm,width=\linewidth,angle=-90}}
\end{center}

\caption[]{Simulated linear polarization histograms: (a) a conal
double profile modeled on B0525+21, and (b) a conal single
(``drifter'') profile modeled on P0820+02.  See text for details.  No
effort has been made to model Stokes $V$ (dotted line). Pulsar names
have been given as ``0000+00'', just to indicate that the profiles are
`simulated'.}
\label{fig:simpol}
\end{figure}

As discussed earlier, conal component pairs exhibit large fractional
linear polarization on their inner edges and pronounced (often nearly
complete) depolarization on their outer edges. The three-way
mode-segregation method provides some vital clues to understanding
this phenomenon. The power corresponding to the weaker SPM is
sufficient to dominate the PPM only on the outer ``wings'' of the
profile. 


The mode-segregation analyses above reveal two important
characteristics of the emission beam configuration.  First, the SPM
emission is generally shifted further outward, away from the magnetic
axis, than the PPM emission.  If this modal radiation is emitted (in
some average sense) by conal beams, then the emission conal region
corresponding to the SPM beam must have a little larger radius than
that of the PPM.

Second, as we saw in Fig.~\ref{fig4} the PPM and SPM power is
substantially out of phase.  Given the small $|\beta|/\rho$ for
1237+25---such that the sightline cuts the conal beams close to the
magnetic axis---the phase difference suggests that emission elements
within the respective modal beams are offset in magnetic azimuth! 
And, indeed, this is just the polarized-beam configuration observed in
the rotating subbeam systems of conal single {\bf S$_{\rm d}$} pulsars
B0809+74 (Rankin \etal\ 2002) and 0943+10, where systematic longitude
offsets between the modes (at $|\beta|/\rho \sim 1$) also indicate
offsets in magnetic azimuth.  In summary, the modal conal emission
patterns are offset in both magnetic colatitude and azimuth.

We can begin to conceive, given the above observational indications,
how complex are the modal depolarization dynamics of conal beams.  The
familiar polarization properties of conal component pairs are produced
by central sightline trajectories (small $|\beta|/\rho$) and represent
an angular average over the modal ``beamlets''.  For conal single
({\bf S$_{\rm d}$}) stars, however, the impact angle $|\beta|$ is very
close to the radius of the emission cone $\rho$, and the observed
average polarization will depend first on just how the sightline cuts
the modal cones, and second on how this modal power is both angularly
and temporally averaged.

In order to understand this situation more fully, we have attempted to
simulate the depolarization processes in conal single and double
pulsars.  To do so, we generated an artificial pulsar signal such as
would be detected by a pulsar backend connected to a radio telescope
({\it e.g.}, WSRT with its {\tt PuMa} processor).  We computed this
(partially) polarized signal using the recipe given in our Appendix, 
together with a rotating subbeam model interacting with a specific 
observer's sightline.  This subbeam system, with pairs of modal 
``beamlets'' which could be offset in both magnetic colatitude $\rho$ 
and azimuth, flexibly modeled properties seen in both the {\bf
S$_{\rm d}$} and {\bf D} stars; and model pulses sequence were 
computed using relations very much like the inverse cartographic
transform in Deshpande \& Rankin (2001).  Further, a low-level,
non-drifting and unpolarized component, with a Gaussian-shaped 
pattern peaked along the magnetic axis, could be added to simulate 
weak core emission.  Here, we have so far ignored the nature of 
the circular polarization but hope to address it in future work.

The modal ``beamlet'' pairs rotate rigidly with a period $\hat{P}_3$
around the magnetic axis, with their rotation phase ``locked'' to each
other. Their respective trajectories have different radii (offset in
magnetic azimuth), and the ``beamlets'' also have somewhat different
radial widths. These characteristics are required in order both to 
permit high polarization on the inner edges of conal component pairs 
and to ensure that their outer edges are fully depolarized. In order 
to specify the radial illumination pattern of the modal ``beamlets'', 
we have used a hybrid function with ranges of both Gaussian-like and 
exponential behavior,
\begin{equation}
P(\theta)\;=\; \frac{\exp[-\theta/2\sigma^2] + \exp[ -\theta^2/
2\sigma^2]}{1 + \exp[-\theta^2/2\sigma^2]}
\end{equation}
\noindent
where $\theta$ is the radial distance from the centre of the ``beamlet'',
and $\sigma$ is its Gaussian-like {\it rms} scale.  This functional 
form was chosen to provide a smoothly falling function near the 
``beamlet'' peak and an exponential-like behavior on its edges. 
Although there is no physical basis for this choice, it seems to 
reproduce rather nicely the outer edges of the profiles shown in 
Fig. \ref{fig1}. 

A schematic picture of our simulation model is then shown in 
Figure~\ref{fig:simul_beam}. Othogonally polarized sets of modal 
``beamlets'' are shown in greyscale, which slowly rotate so as to 
form the two modal subcones.  The peaks of the respective PPM 
($\parallel$ polarization) and SPM ($\perp$ polarization) subcones 
are indicated by solid and dashed curves.  A weak, non-drifting and 
unpolarized core beam is also included.  

Figure~\ref{fig:simpol} then shows some results from our simulations.
The top panel represents an attempt to model a conal double ({\bf D})
pulsar with properties similar to the canonical pulsar B0525+21.  So,
we have taken $\alpha$, $\beta$, and $P$ some 21$\deg$, 1.5$\deg$ and
3.75 s, respectively. Further, in order to model its 430-MHz profile,
we took the mean radii of its two modal subbeam systems to be 3.0 and
3.6$\deg$.  We also assumed that its two orthogonal modes are fully 
linearly polarised. The rotating-subbeam system corresponding to the 
PPM and SPM each have 8 subbeams, with $\sigma$ scales of 1.3$\deg$ 
and 0.88$\deg$ each, and the peak amplitudes of the SPM ``beamlets'' 
are about 60\% of their PPM counterparts.

We also modeled the central core component [which for B0525+21 should 
have an observed width of 1.77$\deg$ (see Paper IV, eq. 5)] as a 
non-drifting, unpolarized pencil beam with a Gaussian profile centered 
along the magnetic axis.  However, since our sightline intersects this 
weak emission far off on its beam edge ($\beta\sim$ 1.5 degrees), it 
contributes little to the model sequence and profile.

As can then be seen, the fractional linear polarization of the model 
profile reaches a maximum on the inner edges of the two components 
and drops sharply on their outer edges, just as is observed [{\it c.f.}, 
Fig.~\ref{fig2} and, for instance, Blaskiewicz \etal\ (1991)]. Note 
the `S'-shaped PA traverse and the parallel modal PA stripes on their 
outer edges, which correspond to those samples where the SPM sometimes 
dominates the PPM. This modal display is also usefully compared 
directly with the corresponding 430-MHz PA histogram of B0525+21 in 
Hankins, Rankin \& Eilek (2002).  Clearly, we have made no attempt to 
model the circular polarization. 

The bottom panel of Fig.~\ref{fig:simpol} then depicts our effort to
simulate the polarized emission-beam configuration of a conal single
({\bf S$_{\rm d}$}) star, and here we have taken pulsar B0820+02 as 
an example. In this case we took $\alpha$, $\beta$, and $P$ some
19$\deg$, 5.5$\deg$ and 0.865 s, and the radii of the subcones
corresponding to the PPM and SPM were 4.5$\deg$ and 5.1$\deg$,
respectively---nearly equal to $|\beta|$ as expected. In this case, 
the weak unpolarized core beam has a computed with of 4.05$\deg$, 
and again contributes little to the model sequence and profile. 
Of course, we cannot know for {\bf S$_{\rm d}$} stars just how far 
out the sightline crosses the conal beam, so we can adjust this point 
slightly to match particular polarization characteristics.

The ratio of the two subcone radii in the respective examples chosen
above are different.  In the first case (B0525+21), it is 0.83
(3$\deg$/3.6$\deg$), while in the second case (B0820+02) 0.88. 
Although these two ratios are quite close in their values, it is
unclear what might cause this ratio to vary from star to star. By
contrast, within the dynamical picture we present here, the aggregate
polarization properties must be independent of parameters such as
$P_2$ (the subpulse separation in longitude), $P_3$ (the time for
subpulse to drift through a longitudinal interval of $P_2$), and
$\hat{P}_3$ (the subbeam circulation period). It is also important 
to note that the aggregate profile characteristics are completely 
independent of the total number of circulating subbeams.

In two particulars, our simulations depart significantly from what 
is observed: First, as a consequence of assuming that the modal 
emission is fully linearly polarized, we generally obtain higher 
levels of aggregate linear polarization than is seen in the profiles 
we are attempting to model. This suggests, as yet inconclusively, 
that the modal beams are not fully polarized. Second, we find much 
less scatter in the model PAs around the geometrically determined 
PA traverse. While the best observations have for some time suggested
that this excessive scatter could not be the result of the system
noise, more quantitative statements have not been easy to make.
However, McKinnon \& Stinebring (1998; 2000) have developed
statistical analysis tools which should make a more meaningful
assessment practical, and we plan to pursue this question in a 
future paper (Ramachandran \& Rankin 2002).

\section*{Summary and Discussion}

The results of this paper can be summarized succinctly: Conal beams
have a rotating subbeam structure which also entails displacements
between the PPM and SPM radiation in both magnetic latitude and
azimuth.  This results in the outer-edge depolarization seen in conal
component pairs as well as the complex (and often nearly complete)
depolarization found in pulsar profiles that represent an oblique
sightline trajectory along the outer edge of the conal beam.  It also
provides a new and fundamental reason why the modal emission is so
often statistically ``disjoint'' (see Cordes \etal\ 1978). These
characteristics of conal emission can be identified in a variety of
ways, and the conclusions verified by detailed models and simulations.

It is also likely that these effects largely explain the frequency
dependence of the fractional linear polarization in the classic cases
of conal double profiles ({\it i.e.}, B1133+16) first problematized by
Manchester, Taylor \& Huguenin (1973).  Many more recent studies have
pointed to both the the secular decline at high frequencies and the
mid-band ``break'' point below which the aggregate fractional linear
increases no further ({\it e.g.}, McKinnon 1997). And closely
associated with these profile effects are pulse-sequence phenomena
ranging from the purported ``randomizing'' of the PA at high
frequencies to distributions of polarization characteristics in
subpulses.  If we understand that the PPM and SPM ``cones'' have a
significant displacement in magnetic colatitude at meter wavelengths,
then radius-to-frequency mapping (see Paper VII) almost certainly
tends to reduce this displacement at higher frequencies.  Perhaps the
characteristic depolarization of conal beams at very high frequencies
(as well as the ``random'' PAs) is simply the result of modal beam
overlap.  Perhaps the ``breaks'' mark the frequency at which the modal 
beams diverge to the point that no further depolarization occurs. It 
will be satisfying to test these ideas in future detailed studies.

The origin of ``orthogonal mode'' emission has been a topic of debate
for decades. Numerous models have been suggested wherein the two modes
are intrinsic to the emission mechanism itself (e.g., Gangadhara 1997)
and, lacking strong contrary evidence, some bias has developed in
favor of this assumption. However, direct production implies that the
modes be fully (elliptically) polarized and associates them with a
basic emission mechanism which is itself still unknown [for a review,
see Melrose (1995)].

The possibility that disjoint orthogonal modes can arise from
propagation effects was also explored very early by several authors
(Melrose 1979; Allan \& Melrose 1982). The central idea here is that
the natural wave modes, being linearly polarized in two orthogonal
planes, have different refractive indices, and become separated in
space and angle during their propagation. This phenomenon of
refraction in the magnetosphere was explored rigorously by Barnard \&
Arons (1986).

A recent work of Petrova (2001) has addressed these issues in greater
detail.  According to her model, the primary pulsar radiation is
comprised of only one (ordinary) mode, which is later partially
converted into extraordinary-mode emission. It is in this conversion
that the orthogonal polarization modes arise.  Therefore, the
transition from one mode to the other, as observed in pulsar emission,
can be understood as due to switching between a ``significant'' and
``insignificant'' conversion. At any given time and pulse longitude,
the disjoint mode is the sum of two {\it incoherently} superposed
modes. This nicely explains the partial polarization observed in the
pulsar radiation.

Conversion to the extraordinary mode, in Petrova's model, is easiest
for those rays which are refracted outward, away from the magnetic
axis, and such emission apparently comprises the conal beam---though
her work yet gives no understanding about why there should be two 
distinct types of conal beams that are both present in some cases.  
It is further unclear how the ordinary or extraordinary mode would 
be polarized, thus how it then could be identified as a specific PPM 
or SPM in a given pulsar, and why one or the other should experience 
a greater angular offset in magnetic colatitude. Finally, this model 
appears to be fully symmetric in azimuth, so that it is again hard 
to see how the wave-propagation effects can explain the observed 
angular offsets in magnetic azimuth.  

To summarise, the important conclusions of this work are as follows:

\begin{itemize}
\item The average profiles of pulsars with conal component pairs 
exhibit low fractional polarization on their outer edges, and often 
high fractional linear polarization on their inner edges.
\item This very general behavior can be understood in terms of 
the dynamic averaging, along the observer's sightline, of emission 
from a rotating system of subbeams with systematic modal offsets.  
\item The ``beamlet'' pairs corresponding to the PPM and SPM emission 
are offset not only in magnetic latitude, but also in magnetic 
longitude. In other words, the respective average modal beams 
can be visualized as distinct emission cones with somewhat 
different angular radii.  Dynamically, the ``beamlet'' pairs 
maintain a fixed relation to each other as they circulate about 
the magnetic axis.
\item The outer-edge depolarization requires that the PPM and SPM 
subcones have nearly identical specific intensity and angular 
dependence in this region.  This would appear to place strong 
constraints on their physical origin.  
\item The causes of these remarkable angular offsets between the PPM
and SPM emission is unclear. Propagation effects can more easily
explain the shifts in magnetic latitude than longitude.

\end{itemize}



\acknowledgments

We thank Avinash Deshpande for important analytical assistance and
Mark McKinnon, Russell Edwards, and Ben Stappers for discussions 
and critical comments.  We also thank the latter two for help with 
the WSRT observations.  One of us (JMR) also gratefully acknowledges 
grants from the Netherlands Organisatie voor Wetenschappelijk 
Onderzoek and the US National Science Foundation (Grant 99-86754).  
Arecibo Observatory is operated by Cornell University under contract 
to the US NSF.

\section*{Appendix: Simulation of Partially Polarized Radiation}
\label{sec-polsim}
Let us consider the two complex signals,
$X(t)\,=\,[X_R(t)\,+\,j\,X_I(t)]$ and
$Y(t)\,=\,[Y_R(t)\,+\,j\,Y_I(t)]$, which represent the Nyquist-sampled
baseband voltages from the two orthogonal linear dipoles ({\bf X} \&
{\bf Y}) of a radio telescope. The subscripts $R$ and $I$ indicate the
real and the imaginary parts of the complex signal, and
$j\;=\;\sqrt{-1}$. The Stokes parameters are defined as
\begin{eqnarray}
I &=& \langle X\;X^{\star}\;+\;Y\;Y^{\star}\rangle \nonumber\\ 
Q &=& \langle X\;X^{\star}\;-\;Y\;Y^{\star}\rangle \nonumber\\ 
U &=& \langle {\tt Re}[X\;Y^{\star}\;+\;X^{\star}\;Y]\rangle \nonumber\\ 
  &=& \langle 2\;|X|\;|Y|\; \cos\theta(t)\rangle\nonumber\\ 
V &=& \langle {\tt Im}[X\;Y^{\star}\;+\;X^{\star}\;Y]\rangle \nonumber\\ 
  &=& \langle 2\;|X|\;|Y|\; \sin\theta(t)\rangle 
\label{eq:stokes}
\end{eqnarray}
\noindent
The $\langle$ $\rangle$ symbols indicate time-averaging, {\tt Re} 
and {\tt Im} the real and imaginary parts, and superscript 
``$\star$'' the complex conjugate. $Q$ and $U$ describe the 
linear and $V$ the circular polarization, obeying the well known 
inequality, $I \geq \sqrt{Q^2+U^2+V^2}$. The angle $\theta$ is the
phase between $X(t)$ and $Y(t)$, which is given by $\theta\, = \,
[\tan^{-1}(X_I/X_R)\;-\; \tan^{-1}(Y_I/Y_R)]$.

From the sampling theorem, we know that a signal varies at 
a rate given by the reciprocal of the bandwidth $\Delta\nu$; 
samples having this resolution are fully polarized and can 
be represented by a point on the Poin\'care sphere. Polarimetry, 
then, always entails averaging over a time sufficient to 
adequately reduce the statistical errors.

To generate a realistic partially polarized Nyquist-sampled 
baseband signal, we adopted the following procedure. A 
{\it randomly} polarized voltage sample in the {\bf X}-dipole 
was defined as
\begin{eqnarray}
X_{Ru}^i &=& \sqrt{1/2}\;P_r\;G_x^i\;\cos(\phi) \nonumber\\
X_{Iu}^i &=& \sqrt{1/2}\;P_r\;G_x^i\;\sin(\phi) \nonumber\\
{\rm where}\;\;\;\phi &=& 2\pi\;U_x^i \ ,
\end{eqnarray}
\noindent
where $P_r$ is its amplitude, $G_x^i$ is a Gaussian-distributed 
random variable with zero mean and unity {\it rms} amplitude, 
and $U_x^i$ a uniform random variable with equal density between 
0 and 1.  Similarly, for the {\bf Y}-dipole,
\begin{eqnarray}
Y_{Ru}^i &=& \sqrt{1/2}\;P_r\;G_y^i\;\cos(\phi) \nonumber\\
Y_{Iu}^i &=& \sqrt{1/2}\;P_r\;G_y^i\;\sin(\phi) \nonumber\\
{\rm where}\;\;\;\phi &=& 2\pi\;U_y^i  \ ,
\end{eqnarray}
\noindent
and $G_y^i$, $U_y^i$ are different random variables as above.  

For linear polarization, the signal voltages are 
\begin{eqnarray}
X_{Rl}^i &=& \cos\chi\;P_l\;G_l^i\;\cos(\phi) \nonumber\\
X_{Rl}^i &=& \cos\chi\;P_l\;G_l^i\;\sin(\phi) \nonumber\\
Y_{Rl}^i &=& \sin\chi\;P_l\;G_l^i\;\cos(\phi) \nonumber\\
Y_{Rl}^i &=& \sin\chi\;P_l\;G_l^i\;\sin(\phi) \nonumber\\
{\rm where}\;\;\;\phi &=& 2\pi\;U_l^i \ ,
\end{eqnarray}
\noindent

and for circular polarization they are
\begin{eqnarray}
X_{Rc}^i &=& \sqrt{1/2}\;P_c\;G_c^i\;\cos(\phi) \nonumber\\
X_{Ic}^i &=& \sqrt{1/2}\;P_c\;G_c^i\;\sin(\phi) \nonumber\\
Y_{Rc}^i &=& - \;\sqrt{1/2}\;P_c\;G_c^i\;\sin(\phi) \nonumber\\
Y_{Ic}^i &=& + \sqrt{1/2}\;P_c\;G_c^i\;\cos(\phi) \nonumber\\
{\rm where}\;\;\;\phi &=& 2\pi\;U_l^i \ ,
\end{eqnarray}
\noindent
and, $G_l^i$, $U_l^i$, $G_c^i$, $U_c^i$ are other random 
variables as above.

The partially polarized observed voltage corresponding to 
a given sample is then
\begin{eqnarray}
X_R^i &=& \left[ X_{Rr}^i\;+\;X_{Rl}^i\;+\;X_{Rc}^i\right] \nonumber\\
X_I^i &=& \left[ X_{Ir}^i\;+\;X_{Il}^i\;+\;X_{Ic}^i\right] \nonumber\\
Y_R^i &=& \left[ Y_{Rr}^i\;+\;Y_{Rl}^i\;+\;Y_{Rc}^i\right] \nonumber\\
Y_I^i &=& \left[ Y_{Ir}^i\;+\;Y_{Il}^i\;+\;Y_{Ic}^i\right] \ ,
\end{eqnarray}
\noindent
and the Stokes parameters corresponding to this sample 
are computed according to Eqn. \ref{eq:stokes}.  These 
Stokes-parameters samples are averaged over $N$ samples to 
achieve a desired resolution and statistical significance 
in the simulated time series.



\begin{thebibliography}{}
\small
\bibitem[Allan \& Melrose (1982)]{AM82} Allan, M. C., \& Melrose, D. B. 
         1982, Proc. Astron. Soc. Australia, 4, 365.  
\bibitem[Backer (1973)]{bac73} Backer, D. C 1973, \apj, 182, 245
\bibitem[Backer \etal\ (1975)]{bac75} Backer, D. C., Rankin, J. M., \& 
        Campbell, D. B.  1975, \apj, 197, 481.
\bibitem[Barnard \& Arons, J. (1986)]{BA86} Barnard, J. J., \& Arons, J. 
         \apj, 320, 138
\bibitem[Bartel \etal\ (1981)]{bkknpsssw} Bartel, N., Kardeshev, N. S., 
         Kuzmin, A. D., Nikolaev, N. Ya., Popov, M. V., Sieber, W., 
	 Smirnova, T. V., Soglasnov, V. A., \& Wielebinski, R. 1981, 
	 \aap, 93, 85. 
\bibitem[Bartel (1981)]{bar81} Bartel, N. 1981, \aap, 97, 384. 
\bibitem[Bartel \etal\ (1982)]{bmsh} Bartel, N., Morris, D., Sieber, W., 
         \& Hankins, T. H. 1982, \apj, 258, 776
\bibitem[Blaskiewicz \etal\ (1991)]{bcw91} Blaskiewicz, M., Cordes, J. M.,
         \& Wasserman, I. 1991, \apj, 370, 643 (BCW)
\bibitem[Cordes, Rankin \& Backer (1978)]{crb78} Cordes, J. M., 
         Rankin, J. M. \& Backer, D. C. 1978, \apj, 223, 961.
\bibitem[Deshpande \& Rankin (1999)]{dr99} Deshpande, A. A. \& Rankin,
          J. M. 1999, \apj, 524, 1008
\bibitem[Deshpande \& Rankin (2001)]{dr01} Deshpande, A. A. \& Rankin,
          J. M. 2001, \mnras, 322, 438
\bibitem[Everett, \& Weisberg (2001)]{ew01} Everett, J. E. \& 
         Weisberg J. M. 2001, \apj, 553, 341.  
\bibitem[Gangadhara (1997)]{g97} Gangadhara, R. T. 1997, \aap, 327, 155.
\bibitem[Gil \& Lyne (1995)]{gl95} Gil, J. \& Lyne, A. G. 1995, 
         \mnras, 276, L55 
\bibitem[Gould \& Lyne (1998)]{gl98} Gould, D. M. \& Lyne, A. G. 1998, 
         \mnras, 301, 235.
\bibitem[Hankins, Rankin, \& Eilek (2001)]{hre01} Hankins, T. H., 
         Rankin, J. M., \& Eilek, J. A.,  2002, in preparation
\bibitem[von Hoensbroech (1999)]{vH99} von Hoensbroech, A., 1999, 
         Ph.D. Thesis, Max Planck Institut f\"ur Radioastronomie, Bonn
\bibitem[von Hoensbroech \& Xilouris (1997a)]{vHX97a} von Hoensbroech, A.,
         \& Xilouris, K. M. 1997a, \aaps, 126, 121
\bibitem[von Hoensbroech \& Xilouris (1997b)]{vHX97b} von Hoensbroech, A.,
         \& Xilouris, K. M. 1997b, \aap, 324, 981
\bibitem[Manchester (1971)]{M71} Manchester, R. N. 1971, \apjs, 23, 283.  
\bibitem[Manchester, Taylor \& Huguenin (1973)]{MTH73} Manchester, R. N., 
         Taylor, J. H., \& Huguenin, G. R. 1973, \apj, 179, L7.
\bibitem[Melrose (1979)]{Mel79} Melrose, D. B. 1979, Australian Journal
         of Physics, 32, 61.
\bibitem[Melrose (1995)]{Mel95} Melrose, D. B. 1995, J. Astrophys. \& 
         Astronomy, 16, 137	 
\bibitem[Manchester et al.(1975)]{man75} Manchester, R. N., Taylor, J. H., 
        \& Huguenin, G. R. 1975, \apj, 196, 83.  
\bibitem[McKinnon (1997)]{mm97} McKinnon, M. M. 1997 \apj, 475, 763.  
\bibitem[McKinnon, \& Stinebring (1998)]{ms98} McKinnon, M. M., \& 
         Stinebring, D. R. 1998 \apj, 502, 883.   
\bibitem[McKinnon, \& Stinebring (2000)]{ms00} McKinnon, M. M., \&
         Stinebring, D. R. 2000 \apj, 529, 435.
\bibitem[Mitra (1999)]{m99} Mitra, D. 1999, Ph.D. Thesis, Raman Research 
         Institute, Bangalore
\bibitem[Mitra \& Rankin (2002)]{mr01} Mitra, D., \& Rankin, J. M. 2002,
         \apj, in press (Paper VII)
\bibitem[Morris \etal\ (1981)]{mgsbt} Morris, D., Graham, D. A., Sieber, W., 
         Bartel, N., \& Thomasson, P. 1981, \aaps, 46, 42.  
\bibitem[Petrova (2001)]{pet01} Petrova, S. A. 2001, \aap, 378, 883.  
\bibitem[Radhakrishnan \& Cooke (1969)]{rad69} Radhakrishnan, V., \& 
         Cooke, D. J. 1969, \aplett, 3, 225
\bibitem[Radhakrishnan \& Rankin(1990)]{r+r90} Radhakrishnan, V. \& 
         Rankin, J. M. 1990, \apj, 352, 258 (Paper V)
\bibitem[Ramachandran et al. (2002)]{} Ramachandran, R., Rankin, J. M., 
         Stappers, B. W., Kouwenhoven, M. L. A., \& van Leeuwen, A. G. J. 
         2002, \aap, 381, 993	 
\bibitem[Rankin(1983a)]{jra83a} Rankin, J. M. 1983a, \apj, 274, 333 (Paper I)
\bibitem[Rankin(1983b)]{jra83b} Rankin, J. M. 1983b, \apj, 274, 359 (Paper II)
\bibitem[Rankin(1986)]{jra86} Rankin, J. M. 1986, \apj, 301, 901 (Paper III)	 
\bibitem[Rankin(1990)]{jra90} Rankin, J. M. 1990, \apj, 352, 247 (Paper IV)
\bibitem[Rankin(1993a)]{jr93a} Rankin, J. M. 1993a, \apj, 405, 285 (Paper VIa)
\bibitem[Rankin(1993b)]{jr93b} Rankin, J. M. 1993b, \apjs, 85, 145 (Paper VIb)
\bibitem[Rankin, Ramachandran, \& van Leeuwen (2002)]{} Rankin, J. M., 
         Ramachandran, R., \& van Leeuwen, A. G. J. 2002, \aap, preprint.
\bibitem[Rankin \& Rathnasree (1997)]{} Rankin, J. M., \& Rathnasree, N. 1997, 
         JAA, 18, 91.
\bibitem[Ruderman \& Sutherland (1975)]{ras75} Ruderman, M. A., \& 
         Sutherland, P. G. 1975, \apj, 196, 51
\bibitem[Ruderman (1976)]{rud76} Ruderman, M. A. 1976, \apj, 203, 206
\bibitem[Sieber \& Oster (1975)]{sao75} Sieber, W., \& Oster, L. 1975, \aap, 38,
         325
\bibitem[Sieber \& Oster (1976)]{oas76} Sieber, W., \& Oster, L. 1976, \apj, 210,
         220
\bibitem[Suleymanova \& Izvekova(1984)]{sul84} Suleymanova, S. A., \& 
         Izvekova, V. A. 1984, Soviet Astronomy, 28, 32
\bibitem[Suleymanova et al. (1998)]{sul98} Suleymanova, S. A., Izvekova, V. A., 
         Rankin, J. M. \& Rathnasree, N. 1998, J. Astrophys. Astron., 19, 1  
\bibitem[Suleymanova \& Pugachev (1998)]{sp98} Suleymanova, S. A., \& 
         Pugachev, V. D. 1998, Astronomy Reports, 42, 252.  
\bibitem[Suleymanova \& Pugachev (2002)]{sp02} Suleymanova, S. A., \& 
         Pugachev, V. D. 2002, Astronomy Reports, 46, 34.  
\bibitem[Taylor \& Cordes (1993)]{tac93} Taylor, J. H. \& 
         Cordes, J. M. 1993, \apj, 401, 674
\bibitem[Taylor \& Huguenin (1971)]{tah71} Taylor, J. H., \& Huguenin, G. R. 
         1971, \apj, 167, 273
\bibitem[Taylor et al.(1971)]{tay71} Taylor, J. H., Huguenin, G. R., 
         Hirsch, R. M., \& Manchester, R. N. 1971, \aplett, 9, 205
\bibitem[Taylor et al. (1993)]{tay93} Taylor, J. H., Manchester, R. N., \& 
         Lyne, A. G.  1993, \apjs, 88, 529 
\bibitem[Weisberg \etal (1999)]{wcld99}
         Weisberg, J. M., Cordes, J. M., Lundgren, S. C., Dawson, B. R., Despotes, J. T., 
         Morgan, J. J., Weitz, K. A., Zink, E. C. \& Backer, D. C., 1999, \apjs, 121, 171
\end{thebibliography}
\end{document}